\documentclass[preprint,12pt]{elsarticle}

\usepackage{bm}
\usepackage[outline]{contour}
\usepackage{dashrule}
\usepackage{chngcntr}
\usepackage{natbib}
\usepackage[T1]{fontenc}
\usepackage{amsmath, amssymb}
\usepackage{graphicx}
\usepackage{booktabs}
\usepackage{array}
\usepackage{float}
\usepackage{multirow}
\usepackage[table,dvipsnames]{xcolor} 
\usepackage{subcaption}
\usepackage[colorlinks]{hyperref}

\AtBeginEnvironment{tabular}{\small}

\newcolumntype{P}[1]{>{\centering\arraybackslash}p{#1}}
\definecolor{lightgray}{gray}{0.9}
\arrayrulecolor{gray} 

\newcommand{\gbox}[1]{\colorbox{YellowGreen!30}{#1}}
\newcommand{\gcell}[1]{\cellcolor{YellowGreen!30}{#1}}

\newcommand{\tbox}[1]{\colorbox{Tan!30}{#1}}
\newcommand{\tcell}[1]{\cellcolor{Tan!30}{#1}}

\begin{document}

\begin{frontmatter}

\title{\textbf{Multimodal Representation Alignment for Cross-modal Information Retrieval}} 

\author[1]{Fan Xu}
\ead{fan.xu@uni.lu}
\ead[orcid]{0009-0008-6246-267X}
\author[1]{Luis A. Leiva\corref{cor1}}
\ead{luis.leiva@uni.lu}
\ead[orcid]{0000-0002-5011-1847}
\cortext[cor1]{Corresponding author.}

\affiliation[1]{
            organization={Department of Computer Science, University of Luxembourg},
            addressline={6, avenue de la Fonte}, 
            city={Esch-sur-Alzette},
            postcode={L-4364}, 
            country={Luxembourg}}

\begin{abstract}
Different machine learning models can represent the same underlying concept in different ways. 
This variability is particularly valuable for in-the-wild multimodal retrieval, where the objective is to identify the corresponding representation in one modality given another modality as input. 
This challenge can be effectively framed as a representation alignment problem. 
For example, given a sentence encoded by a language model, retrieve the most semantically aligned image based on representations produced by an image encoder, or vice versa.
To gain insights into the performance impact of different metrics, embedding spaces, and representation alignment for retrieval tasks, 
we first empirically investigate the geometric relationships between visual and textual embeddings derived from both vision-language models and combined unimodal models. 
We then align these representations using four standard similarity metrics as well as two learned ones, implemented via neural networks of different architectures with varying losses across multiple benchmarks. 
Our experimental findings indicate that cosine similarity consistently outperforms all the investigated metrics in representation alignment tasks, and that Wasserstein distance provides a complementary perspective on cross-modal distributional differences.
We also observe that 
our proposed custom contrastive loss is advantageous over the MSE loss for aligning image and text representations,
for both multilayer perceptrons and transformer-based models. 
Taken together, our findings 
offer novel insights and practical considerations for researchers working in multimodal information retrieval, 
particularly in real-world, cross-modal applications.
Our code is publicly available: \url{https://github.com/StevenXuf/Cross-modal-Feature-Alignment}
\end{abstract}


\end{frontmatter}

\section{Introduction}

Multimodal Deep Learning (MDL) has made significant progress 
thanks to recent advancements in vision and language models.
The Convolutional Neural Network (CNN) architecture~\cite{convolutionalnn},
one of the most outstanding models in Computer Vision,
has recently been superseded by the Transformer architecture~\cite{transformer}, 
in particular the Vision Transformer architecture (ViT)~\cite{vit} when training on very large datasets.
In Natural Language Processing, we can find various Large Language Models (LLMs) 
such as BERT~\cite{bert} 
or the Generative Pre-trained Transformers (GPT) family~\cite{gpt1, gpt2, gpt3}, 
which offer new possibilities for constructing MDL models.

Within the plethora of MDL models, 
Vision-Language Models (VLMs) consisting of two encoders (for images and texts) 
are increasingly being adopted in downstream tasks,
including Information Retrieval (IR), Visual Question Answering (VQA), and image captioning, among others. 
For example, CLIP~\cite{clip} utilizes an image encoder and a text encoder to contrast visual and textual representations, 
achieving competitive zero-shot classification performance.
BLIP~\cite{blip} builds upon the idea of CLIP and uses the Transformer architecture
with a triple loss objective, achieving state-of-the-art (SOTA) performance in some downstream tasks. 
The core of the success of VLMs is attributed to the fact that 
they create representations in a shared embedding space
that accounts for vision and language modalities jointly.
These models have an enormous potential for in-the-wild multimodal IR,
i.e., finding the corresponding representations of one modality given another modality as input. 
This can be framed as a representation alignment task.
For example, given the semantics in a sentence that is encoded by a language model such as BERT, 
find the closest image that conveys the same textual semantics 
but based on the representations provided by an image encoder such as ViT, or vice versa. 
Inspecting whether aligned representations relate to the embedding geometry is of particular importance, as well-aligned representations usually enhance downstream task performance~\cite{modalitygap}. However, it remains unclear whether this improvement is also reflected by the geometry of the embedding spaces, especially when compared to unaligned spaces. Unfortunately, little is known about these phenomena in VLMs and their connection to downstream tasks.

To bridge this gap in the research literature,
we first analyze the latent spaces created by three SOTA VLMs 
and three paired unimodal-model configurations (i.e., the combination of three vision models with three language models). 
Next, we investigate standard metrics to measure cross-modality similarity 
(Euclidean, Manhattan, and Chi-square distances, and cosine similarity)
as well as learned metrics (Mean Squared Error and a custom contrastive loss) 
to align vision and language representations 
extracted by the aforementioned multimodal and combined unimodal models.
We found that asymmetry of retrieval performance exists when aligning representations across multiple public benchmarks, 
and that the learned metrics via neural networks usually underperform cosine similarity by a significant margin. 
We also found that both transformer-based models and Multi-layer Perceptrons (MLPs) effectively align representations regardless of aligned representations from pre-trained VLMs or those of unaligned combined unimodal models when using our custom contrastive objective for cross-modal downstream IR tasks, 
indicating the effectiveness of the custom loss over MSE when capturing cross-modal representation interactions.
We conclude therefore that representation alignment performs best with VLMs that are trained from scratch with contrastive loss.
Our study provides a new perspective for researchers interested in IR using VLMs and representation alignment via neural networks.
Taken together, this paper makes the following contributions:
\begin{itemize}
    \item An exploration of the embedding spaces induced by multimodal models and paired unimodal models, showing that distributional measures such as Wasserstein distance ($W_2$) provide a complementary way to characterize cross-modal differences, while the measured modality gap does not necessarily predict downstream retrieval performance.
    \item A systematic examination of commonly used metrics for measuring cross-modal embedding similarity, 
    including standard and learned ones using different model architectures with both MSE and contrastive losses.
    \item A custom contrastive loss for aligning cross-modal representations, applicable to both pre-trained aligned representations and initially unaligned representations.
    \item A series of comprehensive benchmarking experiments of image-to-text and text-to-image retrieval tasks using various encoder models across multiple datasets.
\end{itemize}
\section{Related Work}

MDL models are typically constructed upon unimodal models, which are designed to process a single modality—such as images, audio, text, or time series data. For the sake of simplicity, this work focuses specifically on VLMs. Consequently, we begin by reviewing foundational models in the vision and language domains, followed by an overview of SOTA VLMs. We then examine contrastive learning methods and their applications to multimodal IR tasks.

\subsection{Vision Foundation Models}

CNN architectures are foundational in the field of computer vision~\cite{cnnsurvey}, serving as a cornerstone of deep learning for a wide range of vision tasks. Among these, ResNet~\cite{resnet} remains a widely adopted architecture due to its robust performance and versatility. More recently, ViT-based models have gained popularity, owing to their inherent parallelization and scalability advantages. Notable examples include ConViT~\cite{convit} for image classification and ViTDet~\cite{vitdet} for object detection.
Concurrently, efforts have been made to enhance CNN performance through architectural innovations. For instance, ConvNeXt~\cite{convnext}, which retains a purely convolutional structure, has demonstrated superior performance over the hierarchical Swin Transformer~\cite{swintransformer} on both object detection and segmentation benchmarks.

Despite the strong performance of modern vision models across a range of tasks, a systematic investigation into the relationship between embedding spaces and downstream retrieval performance remains limited. In this work, we address this gap by analyzing the embedding spaces of various vision and language models.

\subsection{Language Foundation Models}

The Transformer architecture serves as the foundational backbone for all modern LLMs. Built upon this framework, BERT~\cite{bert} introduced deep bidirectional representations learned from unlabeled text; RoBERTa~\cite{roberta} significantly enhanced BERT's performance through more robust training strategies; and DeBERTa~\cite{deberta} further improved pre-training efficiency via a disentangled attention mechanism and an enhanced mask decoder. Taking a different approach, XLNet~\cite{xlnet} employed a permutation-based language modeling objective to unify the strengths of both autoregressive and autoencoding methods.
The GPT family has also seen consistent progress. GPT-1~\cite{gpt1} introduced a single task-agnostic model using generative pre-training followed by discriminative fine-tuning. GPT-2~\cite{gpt2} demonstrated that large-scale language models can acquire a wide range of capabilities without task-specific supervision. GPT-3~\cite{gpt3} showed that scaling model size yields significant gains in zero-shot, one-shot, and few-shot performance, sometimes rivaling fine-tuned models.
Other SOTA LLMs include LLaMA~\cite{llama}, which surpasses GPT-3 in performance despite being an order of magnitude smaller; BLOOM~\cite{bloom}, which achieves strong results via multi-task fine-tuning; and GLM~\cite{glm}, which outperforms both BERT and GPT-style models in various benchmarks.

When combined with vision models, these LLMs allow for a more nuanced exploration of the geometry of multimodal embedding spaces. This enables comparison between jointly trained VLMs and those trained separately, offering insights into cross-modal representation learning.

\subsection{Vision-Language Models}

The proliferation of vision and language models has significantly expanded opportunities for developing novel VLMs. Among these, CLIP~\cite{clip} represents a widely adopted framework, leveraging contrastive loss as its primary training objective. Building upon this paradigm, BLIP~\cite{blip} extends functionality beyond image classification to VQA and retrieval tasks, while BLIP-2~\cite{blip2} advances efficiency by integrating a frozen image encoder with a language model to bootstrap vision-language pretraining.
Concurrently, alternative architectures have emerged: Meta-Transformer~\cite{metatransformer} introduces a unified framework for multimodal representation learning, encoding up to 12 modalities within a shared embedding space without paired data. Similarly, CoDi~\cite{codi} enables parallel generation of arbitrary output combinations from diverse input modalities through representation alignment in diffusion models. Further contributions from VLMs~\cite{flamingo, florence, florence2, visualchatgpt, cogvlm} and multimodal LLMs~\cite{xllm, visualbert, speechgpt, nextgpt} continue to enrich the architectural diversity and training methodologies within VLMs.

However, prior research lacks a systematic comparative analysis of embedding spaces across diverse VLMs. To address this gap, we empirically investigate geometric structures of embedding spaces using three VLMs, contrasting them against unimodal baselines.

\subsection{Contrastive Learning}

The success of VLMs hinges on contrastive learning~\cite{vlmsurvey}, a paradigm initially developed to learn invariant mappings in single-modality datasets. Early contrastive methods like~\cite{contrastivelearning} paired data points using prior knowledge (e.g., class labels) and measured similarity via Euclidean distance between representations in a "vanilla" contrastive loss framework.
This foundation evolved with CPC~\cite{infonce}, which introduced InfoNCE, a contrastive loss based on unnormalized cross-entropy using cosine similarity. InfoNCE maximizes mutual information between latent representations by treating the task as a classification problem over positive and negative pairs. Building on this, MoCo~\cite{moco} proposed a momentum-based framework: a query encoder is trained via backpropagation, while a key encoder is updated via exponential moving averages (EMA). MoCo maintains a dynamic queue of encoded keys from previous batches, enabling efficient sampling of negatives for InfoNCE. Crucially, gradients only update the query encoder, decoupling training stability from batch size.
SimCLR~\cite{simclr} further simplified contrastive learning by relying solely on in-batch negatives. It replaced Euclidean distance with cosine similarity in a normalized, temperature-scaled InfoNCE loss, emphasizing the importance of strong data augmentations. Unlike MoCo, SimCLR requires large batches to sample sufficient negatives but achieves competitive performance with minimal architectural complexity.
Diverging from explicit negative sampling, BYOL~\cite{byol} introduced a bootstrapping approach. An online network learns to predict the output of a slowly evolving target network (updated via EMA) using mean squared error. By eliminating negative pairs entirely, BYOL avoids collapse through asymmetric architectures and momentum updates, demonstrating that contrastive learning can thrive without explicit negatives.

To bridge vision and language modalities, models like CLIP and ALIGN~\cite{align} extended contrastive learning to multimodal data. Both employ InfoNCE with softmax-normalized cosine similarity to align image-text pairs, but differ in scale and data strategy. CLIP uses curated datasets, while ALIGN leverages larger noisy web-scale datasets.
Beyond softmax-based losses, SigLIP~\cite{siglip} introduced a sigmoid loss for image-text contrastive learning using cosine similarity. This approach replaces pairwise softmax normalization with element-wise sigmoid function, enabling symmetry between image-to-text and text-to-image tasks. SigLIP reduces memory overhead by avoiding batch-wise dependencies, achieving efficiency without sacrificing performance.

Nevertheless, contrastive learning promotes a modality gap, causing a narrow cone effect where all representations reside in a small region of high-dimensional spaces, as evidenced by \citet{modalitygap}.
Their work demonstrates that this modality gap substantially impacts downstream performance and is modulated by the loss function’s temperature parameter, concluding that an optimal gap magnitude varies across tasks, benefiting some while hindering others.
Corroborating this, research in recommender systems~\cite{Yilma23b_varecsys} associated a reduced modality gap with enhanced downstream performance.
\citet{contextualembed} also noted that word embeddings from LLMs are highly anisotropic using self-similarity and intra-sentence similarity metrics, indicating a positive cosine similarity even when they are semantically different.

In contrast to prior work, we study neural networks with different architectures using a custom contrastive loss function to directly align cross-modal representations, thereby investigating the viability and efficacy of this approach and providing useful insights.

\subsection{Applications in Information Retrieval}

VLMs facilitate cross-modal IR through similarity-based metrics, proving particularly advantageous for real-world IR applications. Specifically, backbone encoders from multimodal architectures can extract representations from disparate datasets, subsequently identifying latent semantic commonalities via representation alignment. This process involves, for any given representation in one modality, ranking candidate representations across other modalities using a (dis)similarity metric and returning the top-$K$ matches.

Beyond the direct use of VLMs for aligned-representation retrieval, several methodologies enhance cross-modal alignment. \citet{dlretrieval} developed mapping functions to project diverse modalities into a unified embedding space, conceptually analogous to Meta-Transformer frameworks. \citet{crossmodalretrieval} introduced a multi-task learning approach for image-text retrieval that demonstrates empirical superiority over competing methods. Additionally, \citet{multimodaltransfer} employed transfer learning with modality-specific neural networks to optimize cross-modal retrieval performance.
In e-commerce domains, \citet{commerceretrieval} leveraged contrastive loss within unimodal models to learn discriminative representations, retrieving top-$K$ similar products through approximate nearest-neighbor search. Notably, \citet{diffhyper} aggregated hierarchical representations from diffusion model U-nets to establish semantic correspondences across images. Concurrently, \citet{dme} adapted VQA architectures to fuse visual-textual representations for enhanced search capabilities, while \citet{cllv} addressed the modality gap~\cite{modalitygap} via structured attention networks for language-to-vision retrieval.

Despite these methodological advances, the relationship between structural properties of multimodal embedding spaces and retrieval efficacy remains inadequately characterized in the existing literature.

\section{Methodology}
\label{sec:method}

We are interested in aligning learned representations (a.k.a., embeddings, or feature vectors) in different modalities
coming from vision models, language models, or VLMs.
Let $\mathcal{D} = \{(\boldsymbol{x}_{i}, \boldsymbol{y}_{i}), i=1,\dots,N\}$
be a dataset that contains $N$ paired samples consisting of text
$\boldsymbol{x}_i$ and corresponding image $\boldsymbol{y}_i$.
Let $f(\cdot)$ and $g(\cdot)$ be text and image encoders, respectively,
which can either be multimodal or unimodal models. 
The extracted text representations $\hat{\boldsymbol{x}}_{i}=f(\boldsymbol{x}_i)\in\mathbf{P}$ 
and image representations $\hat{\boldsymbol{y}}_{i}=g(\boldsymbol{y}_i)\in\mathbf{Q}$ 
are mapped into a common Euclidean embedding space, 
where $\mathbf{P}\subset\mathbb{R}^n$ is the text embedding space
and $\mathbf{Q}\subset\mathbb{R}^n$ is the image embedding space.

After extracting representations by these models, we map all representations to a 2D space using non-linear dimensionality reduction methods (we tested t-SNE~\cite{tsne} and UMAP~\cite{umap}), and then investigate the geometric structure of embedding spaces and its connection to the performance of retrieval tasks.

To measure the similarity between two representations, 
we can apply multiple metrics to obtain $d(\hat{\boldsymbol{x}}_i,\hat{\boldsymbol{y}}_i)$, 
where $d(\cdot,\cdot)$ is a (dis)similarity function. 
We first use 4 standard metrics (Euclidean, Manhattan, and Chi-Square distance, and cosine similarity; see  \autoref{tab:metrics}) 
that can measure how close or different
two representations $\hat{\boldsymbol{x}}_i\in\mathbb{R}^n$ and $\hat{\boldsymbol{y}}_i\in\mathbb{R}^n$ are, 
motivated by previous work~\cite{hellinger, comparing, efficientretrieval, documentsimilarity}. 
We should note that cosine similarity is in the range of $[-1, 1]$, 
while the dissimilarity metrics are unbounded.
We then find the corresponding representations between modalities
by ranking the values according to each metric and retain the top-$K$ matches.
In other words, for each representation in one modality, 
we compute its similarity against all the representations in the other modality in a dataset
and retrieve the top ranked similarity scores.
We follow this process in both text-to-image and image-to-text retrieval experiments (\autoref{sec:perf_experiments}).

\begin{table}[htbp]
    \renewcommand{\arraystretch}{1.5}
    \centering
    \caption{Standard metrics considered. $\boldsymbol{p}\in\mathbb{R}^n$ and $\boldsymbol{q}\in\mathbb{R}^n$ are representations and $d(\cdot,\cdot)$ is the (dis)similarity between them.}
    \begin{tabular}{ll}
    \toprule
     Euclidean distance& $d(\boldsymbol{p},\boldsymbol{q})=\sqrt{\sum\limits_{i=1}^{n}(p_{i}-q_{i})^{2}}$  \\
     Cosine similarity& $d(\boldsymbol{p},\boldsymbol{q})=\frac{\sum\limits_{i=1}^{n}p_{i}q_{i}}{\sqrt{\sum\limits_{i=1}^{n}p_{i}^{2}}\sqrt{\sum\limits_{i=1}^{n}q_{i}^{2}}}$ \\
     Manhattan distance& $d(\boldsymbol{p},\boldsymbol{q})=\sum\limits_{i=1}^{n}\left|p_{i}-q_{i}\right|$ \\
     Chi-square distance& $d(\boldsymbol{p},\boldsymbol{q})=\frac{1}{2}\sum\limits_{i=1}^{n}\frac{(p_{i}-q_{i})^{2}}{p_{i}+q_{i}}$\\
     \bottomrule
    \end{tabular}
    \label{tab:metrics}
\end{table}

As a step forward, we learn to align representations extracted by different models using either MLPs or cross-attention-based transformer models, to compare the effectiveness of more sophisticated networks against the traditional MLPs.
All the neural networks take as inputs both text and image representations, and output a single scalar in range $[-1, 1]$ through \texttt{tanh} activation function at the final layer, akin to the cosine similarity.
Specifically, we employ only one cross-attention block of hidden dimension 128 and 4 cross-attention heads for the transformer models with \texttt{GELU} activation, and then project the cross-attended embeddings via a fully connected network with two linear layers activated by \texttt{ReLU}. 
For MLPs, we use a fully connected network with two hidden layers, where the first layer and the second layer have 512 and 256 neurons followed by a \texttt{ReLU} activation function, respectively.
We apply dropout of probability 0.1 for both architectures, to avoid overfitting.

We first use the Mean Squared Error (MSE) as the loss function to learn cosine similarity via both MLPs and transformer-based models, 
as MSE is simple yet powerful in regression tasks. 
For the MSE loss, the cosine between image and text representations is used as the targets, while the predictions are learned through neural networks.
This way, the networks are imitating the behaviors of cosine similarity.
We also propose a custom contrastive loss as the objective to learn from representations without relying on the cross-modal cosine similarity, which could cause noisy guidance, especially for unaligned representations.

Rather, the custom contrastive loss directly pushes similar representations together (towards value 1) and compels dissimilar representations apart (towards value -1),
as shown in \autoref{eq:contrastive}:
\begin{equation}\label{eq:contrastive}
    \mathcal{L}(\hat{\boldsymbol{x}}_{i},\hat{\boldsymbol{y}}_{j})=I_{ij}\frac{1}{2}\left(1-\hat{d}_{ij}\right)^{2} + (1-I_{ij})\frac{1}{2}\left(\hat{d}_{ij} + 1\right)^{2}
\end{equation}
where
$\hat{d}_{ij}=\Psi(\hat{\boldsymbol{x}}_{i},\hat{\boldsymbol{y}}_{j})$ 
is the output of the neural network that takes image and text representations as inputs, 
and $I_{ij}$ is the indicator function shown in \autoref{eq:indicator}:
\begin{equation}\label{eq:indicator}
I_{ij} = \begin{cases}
1, & i=j\\
0, & i\neq j
\end{cases}.
\end{equation}
Within a batch $\mathcal{B}$, averaging over the loss $\mathcal{L}(\hat{\boldsymbol{x}}_{j},\hat{\boldsymbol{y}}_{j})$ 
of all matched text representations $\hat{\boldsymbol{x}}_j$ and image representations $\hat{\boldsymbol{y}}_j$ for $j \in \{1,\dots,|\mathcal{B}|\}$, we get the positive loss for matched pairs:
\begin{equation}\label{eq:positve_loss}
    \mathcal{L}_{\mathrm{positive}}\left(\mathcal{B}\right)=\frac{1}{|\mathcal{B}|}\sum\limits_{j=1}^{|\mathcal{B}|}\mathcal{L}(\hat{\boldsymbol{x}}_{j},\hat{\boldsymbol{y}}_{j}),
\end{equation}
while the negative loss is the mean over all unmatched pairs within the batch:
\begin{equation}\label{eq:negative_loss}
    \mathcal{L}_{\mathrm{negative}}(\mathcal{B})
    =
    \frac{1}{|\mathcal{B}|\cdot(|\mathcal{B}|-1)}
    \sum_{i=1}^{|\mathcal{B}|}
    \sum_{\substack{j=1 \\ j\neq i}}^{|\mathcal{B}|}
    \mathcal{L}(\hat{\boldsymbol{x}}_{i},\hat{\boldsymbol{y}}_{j}),
\end{equation}
where $|\mathcal{B}|$ is the total number of samples in the batch. 
The final objective over all data samples in the batch is:
\begin{equation}
    \mathcal{L}_{\mathrm{total}}\left(\mathcal{B}\right)=\frac{1}{2}\left(\mathcal{L}_{\mathrm{positive}}\left(\mathcal{B}\right)+\mathcal{L}_{\mathrm{negative}}\left(\mathcal{B}\right)\right),
\end{equation}
which prevents the negative loss from dominating the training process, thereby avoiding degraded performance.
With our proposed custom contrastive loss function, trained models can predict discriminative scores for pairs.
To ensure consistency of model training and fairness when comparing model performance, all models are trained with a batch size of 64 via Adam optimizer ($\beta_1=0.9, \beta_2=0.999$) 
with a cosine annealing scheduler with an initial learning rate of $5 \times 10^{-5}$ for 100 epochs.
All models are trained using the same configuration across different datasets and loss functions, 
both for text-to-image and image-to-text retrieval tasks, resulting in 144 trained models in total.
More model training details can be found in our code repository.
\section{Experiments}

\begin{figure}[htbp]
    \centering
    \includegraphics[width=\linewidth,keepaspectratio=true]{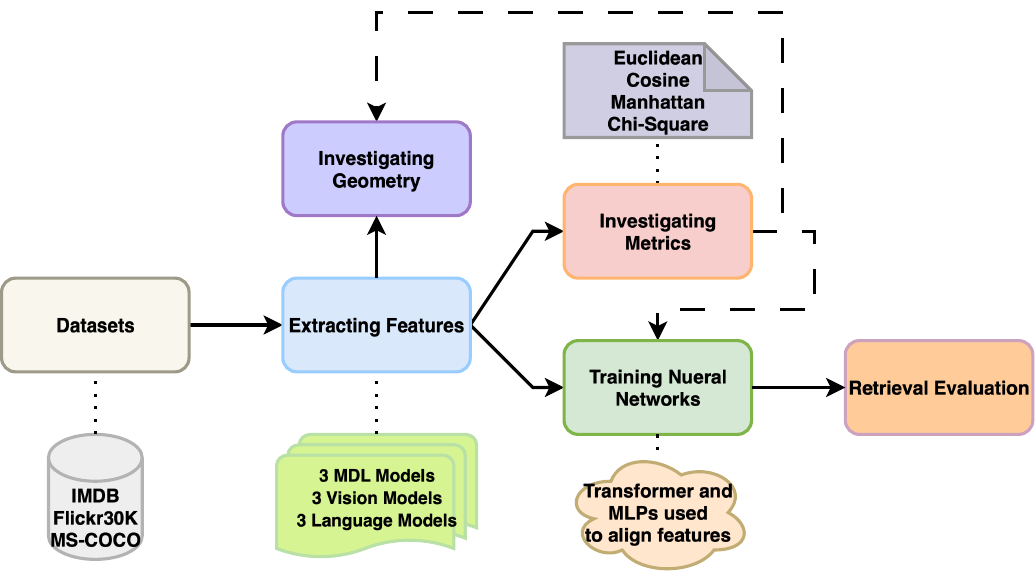}
    \caption{Pipeline of our experiments.}
    \label{fig:pipeline}
\end{figure}

\autoref{fig:pipeline} illustrates our comprehensive experimental pipeline. Initially, we employ both multimodal and unimodal architectures as feature extractors to derive representations from three distinct datasets. Subsequently, we examine the relationship between retrieval performance and the geometric structures of these representations. 
We quantitatively evaluate cross-modal retrieval efficacy using the aforementioned metrics, assessing both image and text representations. Following this analysis, we train neural networks on the extracted representations and validate their performance on retrieval tasks. Finally, we present real-world retrieval experiments to provide further validation.

\subsection{Datasets}
\label{sec:datasets}

We use three public benchmarks: the ``IMDB Vision and NLP'' dataset~\cite{imdb}, which contains 4K movie posters with corresponding names, Flickr30K~\cite{flickr30k}, having 31K images and 155K comments, and MS-COCO~\cite{mscoco} (Train2017 subset), with 118K images and 590K captions.
All datasets are complementary and of different nature, 
from which models' capability of understanding visually different scenarios can be thoroughly tested.
IMDB mainly focuses on visual impression of movie posters, 
Flickr30K contains images showing complex interactions between people and objects, 
and MS-COCO emphasizes multiple objects in context.
To simplify the problem, we assume a one-to-one mapping between modalities,
i.e., one textual description can have only one ground-truth image, and vice versa.
Since Flickr30K and MS-COCO have 5 text captions associated with each image, 
we choose the first description for our experiments.
We resize all images to the same size of $224 \times 224$\,px 
and normalize each pixel value in the range $[0, 1]$ with encoder-specific means and standard deviations, as required by respective model. 
Note that resizing images to $224 \times 224$\,px 
does not necessarily hinder a model's visual understanding capability, 
as evidenced by \citet{sparc}, from which images were resized to that size
while still capturing fine-grained details in images.
Since the captions in all datasets are already well formatted, we process them directly using the tokenizer corresponding to each encoder, without applying any additional preprocessing.

\subsection{Models}
\label{subsec:models}

As mentioned previously, we focus on vision and language modalities, 
and therefore we take representations from separate vision models and language models or use already existing VLMs.
Based on the official naming conventions,
we select three VLMs for their reported SOTA performance: 
\texttt{CLIP-ViT-B/32} (CLIP), 
\texttt{BLIP-base-retrieval-CoCo} (BLIP), and  
\texttt{Meta-Transformer-B/16} (Meta-Transformer).
Moreover, we combine three vision models with three language models:
\texttt{ResNet-34} with \texttt{BERT-small} (ResNet/BERT), 
\texttt{ConViT-base} with \texttt{RoBERTa-base} (ConViT/RoBERTa), 
as well as \texttt{ConvNeXt-small-1K} with \texttt{XLNet-base-cased} (ConvNeXt/XLNet). 
For pre-trained VLMs, we use representations from each official head, while for combined models
we use representations from the layer before the final output layer from the respective unimodal models, 
which are deemed as learned representations. 
We use \texttt{PyTorch} to implement all experiments on three NVIDIA A6000 GPUs (48\,GB of memory).

Vision and language representations have compatible dimensions if they are extracted by any of the 3 VLMs. 
This is natural for VLMs, as they output representations with the same dimensionality. 
However, we intentionally select unimodal models so that paired models can have compatible image and text representations. 
Note that the representation dimensions might be different when using different VLMs as well 
(e.g., representations extracted by CLIP have 512 dimensions 
while those extracted by Meta-Transformer have 768 dimensions).
This, however, does not impact the training of neural networks, 
as their sizes fit the sizes of the individual model input sizes. 
For later analyses in \autoref{sec:embed_space} to \autoref{sec:perf_experiments}, 
all the experiments are conducted on representations extracted 
by either VLMs or combined paired models as mentioned in this section.

\subsection{Embedding Space Analysis}\label{sec:embed_space}

\begin{figure}[htbp]
    \centering
    \includegraphics[width=\textwidth]{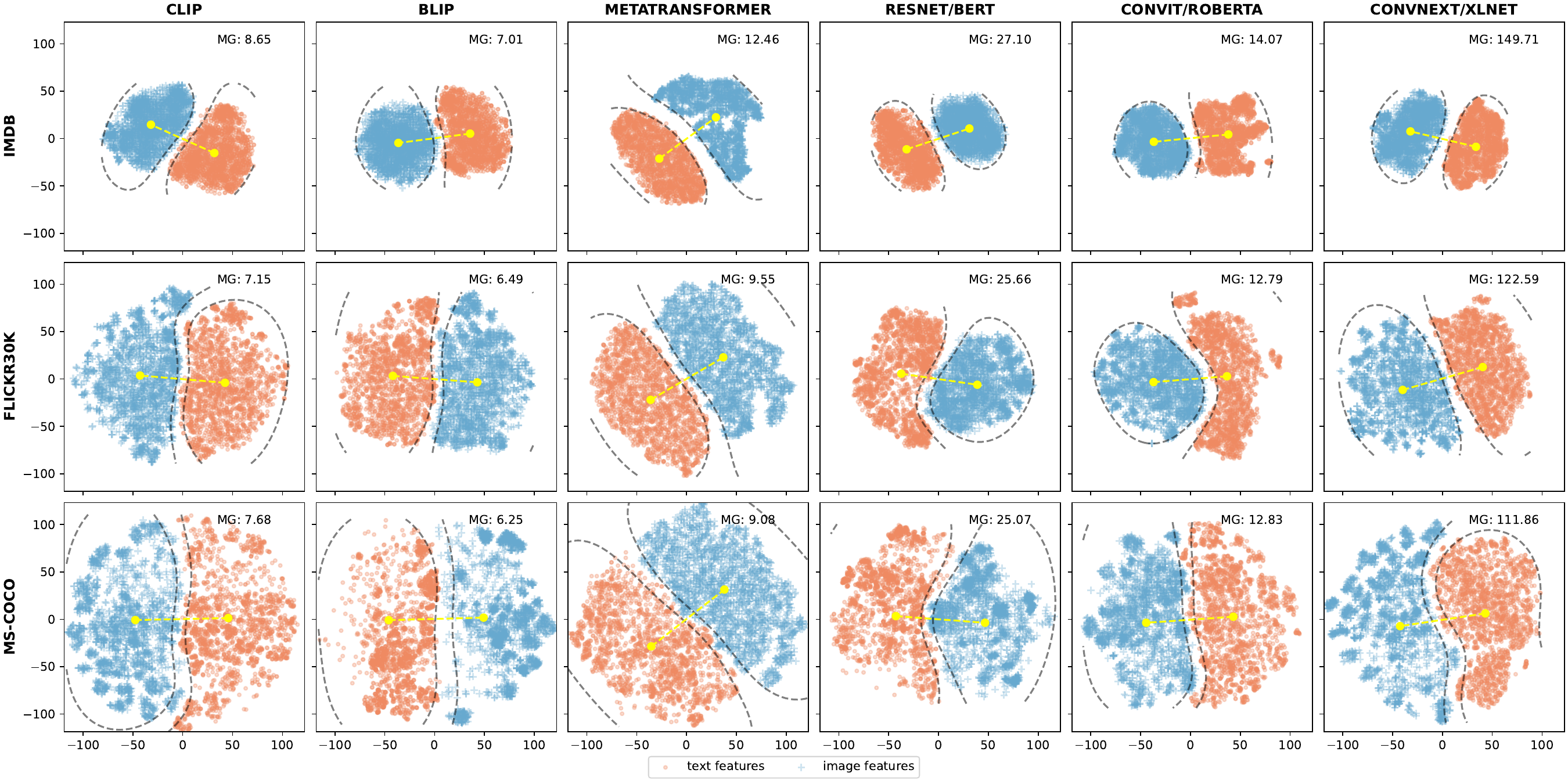}
    \caption{Embedding spaces of VLMs and combined unimodal models on different datasets through t-SNE. \textcolor{NavyBlue}{Image representations} are in blue while \textcolor{BurntOrange}{text representations} are in orange. Dashed lines (\hdashrule[0.5ex][x]{.02\linewidth}{0.5pt}{1mm .2mm}) denote cluster contours, and yellow dots (\textcolor{yellow}{\textbullet}) denote centroids of clusters, for which the modality gap (MG) is shown on the top-right part of each plot.}
    \label{fig:tsne_embeddings}
\end{figure}

We first project the extracted representations for each model and dataset into a 2D space for intuitive visualization.
We use the non-linear projection algorithm t-SNE,
given that linear alternatives such as Principal Component Analysis~\cite{pca} (PCA) 
do not capture well the complex structure of embedding spaces.
UMAP~\cite{umap} also failed to preserve the geometric structure of these embedding spaces.
To ease visualization, the plots in \autoref{fig:tsne_embeddings}
use a random selection of 2K data samples from each dataset, 
to avoid showing over-dense and overlapping clusters.
Dashed lines denote the cluster contours, 
computed by a Support Vector Machine (SVM)~\cite{svm} with a Gaussian kernel.
Yellow dots in \autoref{fig:tsne_embeddings} denote the centroids of each modality cluster. 
The modality gap is computed via Euclidean distance between centroids in the original dimension (instead of the reduced dimension by t-SNE), as defined in \cite{modalitygap}, 
shown on the top-right corner of each subplot.

In IMDB (first row of \autoref{fig:tsne_embeddings}),
we can see that BLIP promotes the shortest modality gap over 3 VLMs, followed by CLIP.
Interestingly, we also observe that even unaligned embeddings of ConViT/RoBERTa and Meta-Transformer
have small modality gaps, except for ConvNeXt/XLNet.
Notably, we observe a large cluster contour separation
in Meta-Transformer among VLMs, 
while the combined unimodal models have similar contour separation.
In Flickr30K (second row of \autoref{fig:tsne_embeddings}), 
the modality gaps for all VLMs are quite similar, 
even though BLIP turns out to be the smallest one. 
No significant differences between cluster contours were observed.
Among all combined unimodal models, 
ConViT/RoBERTa promotes the shortest modality gap, 
and all combined unimodal models have larger modality gaps 
than those of any VLMs.
In MS-COCO (third row of \autoref{fig:tsne_embeddings}), 
BLIP has the smallest modality gap among all VLMs, 
despite that the difference is small overall. 
Again, the modality gaps of all unimodal models are much larger than those of VLMs.
Cluster contours do not exhibit large differences either.
From all plots, the Euclidean distance, used by previous work to measure the modality gap, 
reflects the effects of alignment between varying modalities.

Besides measuring the modality gap between centroids of representation clusters using Euclidean distance, we also compute in \autoref{tab:wasserstein_dist} the Wasserstein-2 distance  to measure the cost of transporting data points from one distribution to another distribution.\footnote{To compute the Wasserstein-2 distance, we randomly split each dataset to 5K batches and average the running distance over all batches to avoid memory issues.}
Note that Wasserstein distance is symmetric with respect to two input distributions.
From \autoref{tab:wasserstein_dist}, we observe the same trend as that of using Euclidean distance to measure the modality gap. 
Concretely, we find that BLIP consistently has the smallest modality gap among other models, 
which serves as an inspiration to align representation distributions between two modalities 
instead of fixed representations extracted by each encoder.
However, models with smaller modality gap do not necessarily show better performance than other models, 
as evidenced by the results in \autoref{sec:perf_experiments}.
For example, although Meta-Transformer shows a closer modality gap among other models to those of CLIP and BLIP, the performance is almost the same as other combined unimodal models overall when using the standard metrics, and even worse when using neural networks to learn the alignment for both text-to-image and image-to-text retrieval tasks.

\begin{table}[htbp]
    \centering
    \caption{Wasserstein-2 distance ($\downarrow$) between image representations and text representations in the original dimension. 
    Best results are in boldface.}
    \resizebox{\linewidth}{!}{
    \begin{tabular}{l *6c}
        \toprule
         &\textbf{CLIP}&\textbf{BLIP}&\textbf{Meta-Transf.}&\textbf{ResNet/BERT}&\textbf{ConViT/RoBERTa}&\textbf{ConvNeXt/XLNet}\\
        \midrule
         \textbf{IMDB}&60.74&\textbf{42.72}&125.12&528.40&114.83&11668.23 \\
         \textbf{Flickr30K}&56.40&\textbf{43.18}&80.52&524.39&99.20&8186.30 \\
         \textbf{MS-COCO}&60.41&\textbf{41.89}&75.68&525.61&102.50&6916.52 \\
        \bottomrule
    \end{tabular}
    }
    \label{tab:wasserstein_dist}
\end{table}

We noted an interesting trend, however: 
embedding distributions tend to be sparser 
as dataset size increases, regardless of the model.
This could be explained by the fact that larger datasets introduce greater semantic diversity, causing representations to occupy a broader region of the embedding manifold.  
This effect reduces local embedding density, resulting in the sparser distributions observed in larger datasets.
Overall, we can conclude that 
(1)~the alignment among different modalities cannot be fully characterized by either spatial distances (e.g., Euclidean distance) nor by distribution differences (e.g., Wasserstein distance); and (2) well-aligned representations of CLIP and BLIP and unaligned representations of other models do not show substantially distinctive geometry for the same dataset.


\subsection{Feature Similarity Analysis}\label{sec:similarity}

\begin{figure}[htbp]
    \centering
    \includegraphics[width=\textwidth]{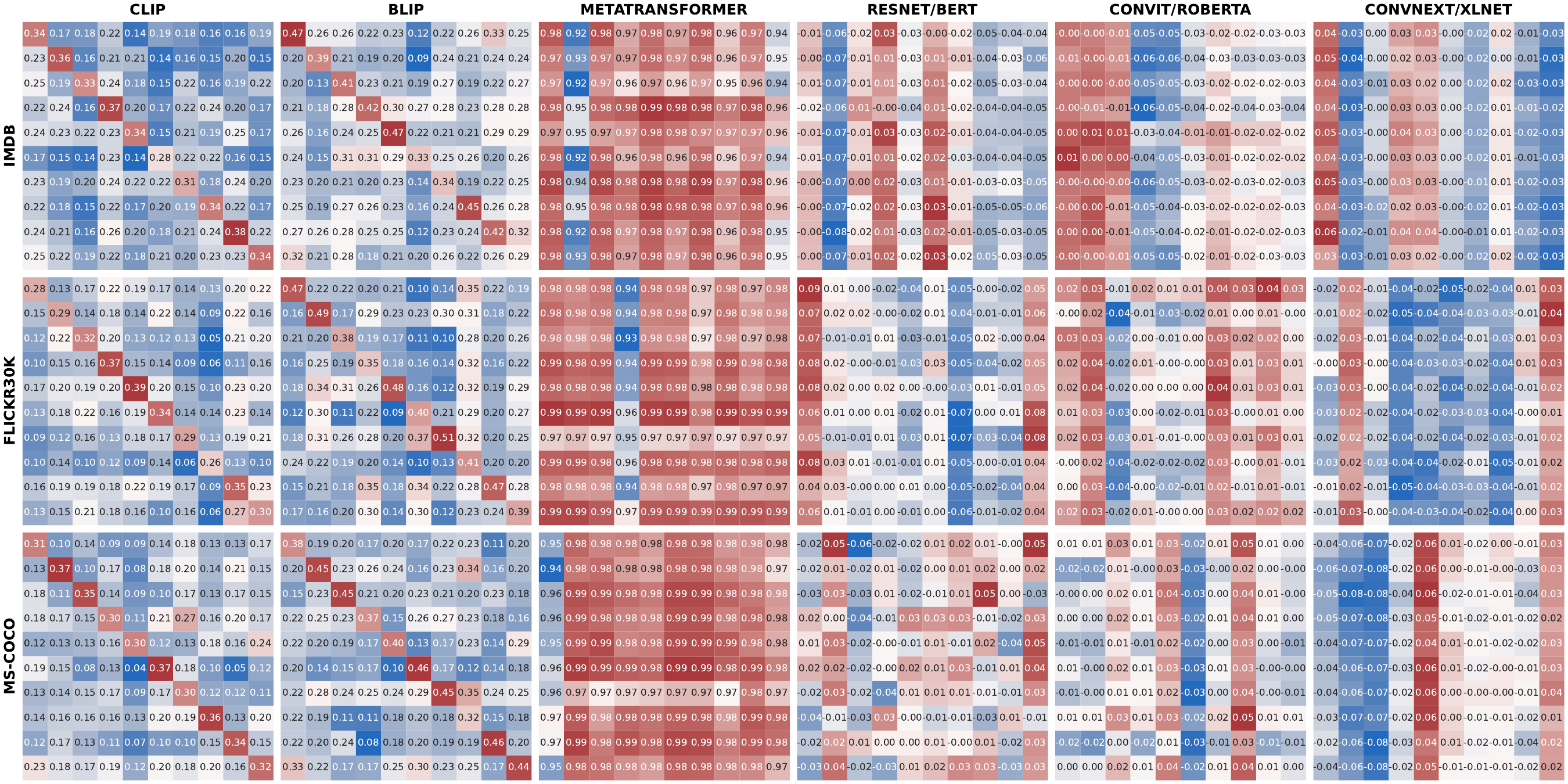}
    \caption{
    Cosine similarity ($\uparrow$) matrix of a random selection of 10 \texttt{<text,image>} pairs. 
    In each cell, cosine similarity is computed based on text 
    and image representation from the dataset showed on the left, extracted by models listed on the top. 
    In each heatmap, the X-axis denotes text representation and the Y-axis denotes the image representation. 
    Red cells have \textcolor{BrickRed}{larger} cosine 
    while blue cells are \textcolor{NavyBlue}{smaller}. 
    As observed, both CLIP and BLIP have clear diagonal trends, whereas other models do not.
    }
    \label{fig:heatmap}
\end{figure}

To corroborate the effect of representation alignment in VLMs and combined unimodal models, \autoref{fig:heatmap} shows the cosine similarity matrices of 10 randomly selected \texttt{<text,image>} pairs (for visualization purposes) from each dataset using different models.
We can see that text representations and corresponding image representations are well aligned by CLIP and BLIP, 
as indicated by the diagonal, 
which is the consequence of harnessing contrastive loss as the whole or a part of the training objectives during pre-training phases. 
In addition, cosine similarity values from VLMs have a higher maximum 
than combined unimodal models overall, 
which suggests that the representation alignment from VLMs 
improves the similarity between representations. 
Note that Meta-Transformer consistently yields cosine similarity more than 0.9 for any text-image pairs, 
which indicates that all representations are clustered in a small region of high-dimensional spaces (its original dimension is 768 while \autoref{fig:tsne_embeddings} shows 2D representations reduced by t-SNE).
This phenomenon imposes challenges when applying it to downstream tasks, 
as all representations are deemed as highly similar across modalities.
Although a retrieval task is performed by ranking similarity scores (cosine similarity in this case), 
the retrieved results will not be much distinguished from each other.
Meta-Transformer is not trained with paired text-image data, so it is not surprising that it cannot pull together similar representations or push apart dissimilar ones.

The extremely high cosine similarities observed in Meta-Transformer indicate that retrieval performance  deteriorates when representations collapse into a highly anisotropic region of the embedding space, forming a smaller narrow cone compared with presentations from other models. 
The experiments of both MLPs and transformer-based models with different losses confirm this phenomenon, since the performance of trained models is lower when compared with other combined unimodal models in \autoref{tab:txt2img} and \autoref{tab:img2txt} in \autoref{sec:perf_experiments}.
This observation suggests that successful multimodal alignment requires a balance between semantic alignment and representation diversity.

Another interesting observation is that the values of cosine similarity 
from all VLMs are always positive,
while those of combined unimodal models are either negative or positive.
This holds for all datasets.
Cosine similarity between any two embeddings extracted by CLIP and BLIP reduces to their dot product, which lies in the range [-1,1].
However, due to the InfoNCE-style contrastive training objective, which maximizes similarity between matched image-text pairs while merely reducing similarity among mismatched pairs, the model is not incentivized to push negative pairs to be orthogonal or negatively aligned.
This leads to an embedding space that is anisotropic, where even semantically unrelated vectors often have positive cosine similarity.
Prior work has observed that embeddings from contrastively trained models tend to cluster within a narrow cone in the shared space \cite{modalitygap,contextualembed}, leading to a high mean pairwise cosine similarity, even among unrelated samples (also happened for Meta-Transformer). 
This behavior reflects a bias in the learned geometry: embeddings are concentrated on a subset of directions within the high-dimensional space, rather than being isotropically distributed.
The narrow cone effect suggests that excessive alignment may reduce embedding diversity and limit discriminability.
This positive similarity bias implies that cosine similarity in CLIP and BLIP does not directly reflect absolute semantic distance. 
Rather, it is better interpreted relatively, e.g., in terms of ranking or within-batch contrast, than as a calibrated measure of dissimilarity.
It is more theoretically sound for the value of cosine similarity to be in the $[-1,1]$ range, 
as similar data points should have a positive correlation 
while dissimilar data points should have a negative correlation. 
Our proposed contrastive loss (\autoref{sec:method}) successfully mitigates this problem: the trained models produce positive scores for matches and negative scores for unmatched pairs. See \ref{app:scores} for and analysis of the predicted scores using our method.

\subsection{Retrieval Performance Analysis}
\label{sec:perf_experiments}

To better understand how representation alignment may affect downstream performance, 
we compute Recall@K in text-to-image and image-to-text retrieval tasks.
Since dataset sizes are quite different, to ensure a fair comparison, 
we use a random subset of 1K \texttt{<image,text>} pairs from each dataset
to compute the 4 common metrics mentioned in \autoref{tab:metrics}.
Then, we use all 4K pairs from IMDB 
and 20K-pair subsets (for efficiency) from both Flickr30K and MS-COCO
to align representations 
through MLPs and transformer-based models with 2 different loss functions,
trained on random disjoint splits 
(80\% of the data for training, 10\% for validation, and 10\% for testing). 
All models are trained with the same splits and the same random seed.
Considering that we use representations of images and texts, the input dimensions of models vary accordingly.
We show the details of representation dimensions and the corresponding model sizes in \autoref{tab:params} in \ref{app:params}.
We trained 144 models in total, corresponding to 2 architectures across 3 datasets along with 6 feature extractors (\autoref{subsec:models})
for both text-to-image and image-to-text tasks,
using either MSE or contrastive loss.

\subsubsection{Text-to-Image Retrieval}

We report the results in \autoref{tab:txt2img}.
Regarding VLMs,
we can see that cosine similarity works best for CLIP and BLIP due to the use of contrastive loss. 
CLIP is the best performer in IMDB, and BLIP is the best performer in Flickr30K and MS-COCO when using cosine similarity. 
Concerning the representation alignment via neural networks,
it turns out that MLPs with custom contrastive loss 
achieve better performance for aligning representations from VLMs compared with MLPs with MSE loss across all the datasets, demonstrating the effectiveness of our proposed loss function.
However, transformer-based models using contrastive loss outperform those using MSE on IMDB, while transformers with contrastive loss slightly fall behind those with MSE loss on Flickr30K and MS-COCO datasets, indicating the potential impact of dataset differences. Both MLPs and transformer-based models underperform cosine similarity,
since learning alignment via neural networks only partially approximate the pre-trained behaviors.

\begin{table}[!t]
    \centering
    \caption{
        Recall ($\uparrow$) at $K\in\{1,5,10\}$ of \emph{text-to-image} retrieval tasks.
        Four commonly used metrics are directly applied to match texts with images without learning, 
        while \tbox{MLPs} and \gbox{transformer} models with either MSE or custom contrastive loss are used to align representations. 
        All experiments used representations extracted either by VLMs or combined unimodal models in \autoref{subsec:models}. Best results on each dataset are in boldface.
    }
    \setlength{\tabcolsep}{0.75em}
    \resizebox{\textwidth}{!}{
    \begin{tabular}{cl|rrr|rrr|rrr|rrr|rrr|rrr}
        \toprule
        & \multicolumn{1}{l}{}\multirow{2}{*}{}&\multicolumn{18}{c}{\textbf{R@1, R@5, R@10} (\%)}\\
        \cmidrule(lr){3-20}
        & \multicolumn{1}{l}{\textbf{Metrics}}&\multicolumn{3}{c}{CLIP}&\multicolumn{3}{c}{BLIP}&\multicolumn{3}{c}{Meta-Transf.}&\multicolumn{3}{c}{ResNet/BERT}&\multicolumn{3}{c}{ConViT/RoBERTa}&\multicolumn{3}{c}{ConvNeXt/XLNet}\\

        \midrule
        \parbox[t]{1em}{\multirow{8}{*}{\rotatebox[origin=c]{90}{\textbf{IMDB}}}}
        & Euclidean      &46.3&63.6&69.2 & 1.1&4.4&6.3 & 0.1&0.3&0.9 & 0.1&0.5&1.0 & 0.1&0.4&1.0 & 0.1&0.6&1.2 \\
        & Cosine         &\textbf{84.4}&\textbf{92.3}&\textbf{94.7} & 77.0&89.5&93.1 & 0.1&0.5&1.1 & 0.3&0.6&1.0 & 0.1&0.6&1.1& 0.2&0.6&0.9 \\
        & Manhattan      &78.1&90.5&92.4 & 0.5&3.2&5.8 & 0.1&0.5&1.1 & 0.1&0.7&1.0 & 0.2&0.6&1.1 & 0.3&0.8&1.0 \\
        & Chi-Square     &0.0&0.1&0.4 & 0.1&0.6&1.4 & 0.0&0.3&1.0 & 0.0&0.8&1.6 & 0.2&0.8&1.5 & 0.0&0.6&0.9 \\
    \cmidrule(lr){2-20}
    & \tcell{MLP (MSE)} &1.0&3.2&6.1&0.2&1.5&2.9&0.2&1.2&2.5&0.2&1.2&2.2&0.2&1.0&1.7&0.0&0.0&0.0 \\
    & \tcell{MLP (Contr.)} &29.0&56.3&65.8&4.4&17.2&26.0&0.0&3.2&6.9&1.0&3.4&7.6&1.5&5.7&11.3&0.5&2.9&5.7 \\
    & \gcell{Transformer (MSE)} &1.2&5.4&9.8&0.7&2.2&3.9&0.5&1.0&2.0&0.2&1.0&2.5&0.2&1.5&2.7&0.2&1.0&2.5 \\
    & \gcell{Transformer (Contr.)} &18.9&44.5&56.0&28.3&54.8&68.6&1.2&3.4&7.6&0.7&6.6&10.6&1.7&3.7&8.8&1.0&3.7&7.1 \\

        \midrule
        \parbox[t]{1em}{\multirow{8}{*}{\rotatebox[origin=c]{90}{\textbf{Flickr30K}}}}
        & Euclidean&16.5& 45.8& 57.2&0.4& 5.1& 10.6&0.1& 0.7& 1.2&0.1& 0.5& 1.1&0.1& 0.5& 1.1&0.1& 0.4& 0.9\\
        & Cosine&67.2& 87.9& 92.4&\textbf{70.6}& \textbf{90.7}& \textbf{94.3}&0.0& 0.6& 1.2&0.1& 0.5& 0.7&0.2& 0.4& 0.7&0.1& 0.5& 1.0\\
        & Manhattan&51.0& 77.7& 85.2&0.5& 5.1& 8.7&0.0& 0.6& 1.3&0.1& 0.4& 0.9&0.1& 0.5& 1.2&0.0& 0.4& 1.2\\
        & Chi-Square&0.1& 0.2& 0.5&0.1& 0.5& 0.7&0.0& 0.1& 0.4&0.0& 0.2& 1.0&0.1& 0.2& 0.9&0.1& 0.4& 0.5\\
    \cmidrule(lr){2-20}
    & \cellcolor{Tan!20}{MLP (MSE)} &33.7&64.1&74.8&15.5&40.9&54.0&0.1&0.3&0.6&0.1&0.2&0.4&0.1&0.3&0.6&0.1&0.3&0.6 \\
    & \cellcolor{Tan!20}{MLP (Contr.)} &40.8&72.7&81.3&23.9&51.4&64.5&0.1&1.0&1.8&2.5&10.6&17.1&5.0&16.0&24.7&1.6&7.8&13.4 \\
    & \gcell{Transformer (MSE)} &35.7&63.7&75.4&40.1&68.6&78.2&0.1&0.3&0.5&0.1&0.3&0.5&0.1&0.4&0.6&0.1&0.3&0.5 \\
    & \gcell{Transformer (Contr.)} &29.7&60.8&73.8&31.7&63.4&75.4&0.3&1.6&2.8&3.9&12.7&20.2&6.2&21.2&31.6&3.8&13.2&20.1 \\

        \midrule
        \parbox[t]{1em}{\multirow{8}{*}{\rotatebox[origin=c]{90}{\textbf{MS-COCO}}}}
        & Euclidean&17.6& 40.0& 50.3&0.5& 4.1& 6.9&0.0& 0.4& 1.1&0.1& 0.5& 0.8&0.2& 0.5& 1.2&0.0& 0.5& 1.1\\
        & Cosine&49.3& 76.4& 88.2&\textbf{67.0}& \textbf{91.2}& \textbf{95.9}&0.0& 0.4& 1.1&0.4& 0.5& 0.9&0.0& 0.6& 0.9&0.1& 0.7& 0.8\\
        & Manhattan&38.8& 68.7& 80.1&1.0& 3.7& 6.1&0.0& 0.4& 1.4&0.1& 0.5& 0.8&0.0& 0.5& 1.1&0.1& 0.6& 1.1\\
        & Chi-Square&0.2& 0.5& 0.9&0.1& 0.6& 0.9&0.0& 0.1& 0.6&0.1& 0.3& 1.1&0.0& 0.2& 0.7&0.1& 0.8& 1.1\\
        \cmidrule(lr){2-20}
    & \tcell{MLP (MSE)} &23.4&52.3&66.2&15.1&43.4&58.3&0.1&0.3&0.5&0.0&0.2&0.4&0.0&0.2&0.7&0.1&0.2&0.4 \\
    & \tcell{MLP (Contr.)} &26.0&57.7&70.3&21.6&51.2&67.6&0.2&1.1&2.1&4.2&14.1&22.3&4.2&16.6&28.0&2.2&10.1&17.7 \\
    & \gcell{Transformer (MSE)} &26.2&55.1&69.1&38.4&70.2&81.9&0.0&0.2&0.5&0.0&0.2&0.4&0.1&0.4&0.6&0.1&0.2&0.6 \\
    & \gcell{Transformer (Contr.)} &22.0&51.7&66.7&31.6&64.4&78.8&0.2&1.7&3.7&3.5&14.4&23.4&5.1&18.8&30.5&2.7&13.1&22.8 \\
        
        \bottomrule
    \end{tabular}
    }
    \label{tab:txt2img}
\end{table}

Concerning the unaligned Meta-Transformer and combined unimodal models,
all performed worse based on the 4 metrics. 
Again, MLPs and transformers with contrastive loss outperform both architectures with MSE loss to align the unaligned representations,
where the transformer with contrastive loss is the best performer among all 4 types of trained models.
Overall, the experiment results demonstrate the effectiveness of our proposed loss function over MSE loss, and transformers are more effective at learning the interactions than MLPs when aligning representations from unaligned vision and language models.

\subsubsection{Image-to-Text Retrieval}

The results are shown in \autoref{tab:img2txt}.
Note that retrieval results using images are not equal to those using texts (non-symmetrical),
since, as shown in \autoref{fig:heatmap}, 
the cross-modality cosine similarity matrices are not symmetric.\footnote{%
In our experiments, given an embedding in one modality, 
we compute the cosine similarity against \emph{all} embeddings in the other modality, and vice versa.
}
We can see that, among all VLMs, 
CLIP outperforms BLIP on IMDB and Flickr30K datasets when using cosine similarity, 
while BLIP performs better than CLIP on MS-COCO for image-to-text retrieval. 
Both CLIP and BLIP outperform Meta-Transformer by large margins, 
whose performance is as poor as those of combined unimodal models.
Similar to previous analysis, the MLPs with contrastive loss outperform MLPs with MSE across all datasets, illustrating the superiority of our proposed contrastive loss over MSE for MLPs.
On the contrary, transformers with contrastive loss outperform on IMDB dataset, while they perform slightly worse than those with MSE loss on Flickr30K and MS-COCO datasets.
This indicates that alignment of representations is affected by both datasets and feature extractors.

\begin{table}[!t]
    \centering
    \caption{
        Recall ($\uparrow$) at $K\in\{1,5,10\}$ of \emph{image-to-text} retrieval tasks.
        Four commonly used metrics are directly applied to match images with texts without learning, 
        while \tbox{MLPs} and \gbox{transformer} models with either MSE or custom contrastive loss are used to align representations. 
        All experiments used representations extracted either by VLMs or combined unimodal models in \autoref{subsec:models}. Best results on each dataset are in boldface.
    }
    \setlength{\tabcolsep}{0.75em}
    \resizebox{\textwidth}{!}{
    \begin{tabular}{cl|rrr|rrr|rrr|rrr|rrr|rrr}
        \toprule
            & \multicolumn{1}{l}{}\multirow{2}*{}&\multicolumn{18}{c}{\textbf{R@1, R@5, R@10} (\%)}\\
            \cmidrule(lr){3-20}
            & \multicolumn{1}{l}{\textbf{Metrics}}&\multicolumn{3}{c}{CLIP}&\multicolumn{3}{c}{BLIP}&\multicolumn{3}{c}{Meta-Transf.}&\multicolumn{3}{c}{ResNet/BERT}&\multicolumn{3}{c}{ConViT/RoBERTa}&\multicolumn{3}{c}{ConvNeXt/XLNet}\\

            \midrule
            \parbox[t]{1em}{\multirow{8}{*}{\rotatebox[origin=c]{90}{\textbf{IMDB}}}}
            & Euclidean&31.0& 46.8& 52.4&63.9& 82.2& 86.6&0.0& 0.5& 1.2&0.1& 0.4& 1.1& 0.4& 1.1&0.1& 0.1& 0.5&1.0\\
            & Cosine&\textbf{83.4}& \textbf{91.4}& \textbf{93.8}&73.4& 88.1& 91.1&0.1& 0.6& 1.2&0.0& 0.5& 1.1&0.1& 0.8& 1.4&0.0& 0.6& 1.7\\
            & Manhattan&79.1& 89.7& 92.5&49.1& 71.0& 78.1&0.1& 0.5& 0.7&0.0& 0.1& 0.6&0.1& 0.5& 1.1&0.1& 0.5& 1.1\\
            & Chi-Square&0.1& 0.1& 0.3&0.0& 0.4& 1.3&0.0& 0.3& 0.3&0.0& 0.8& 1.4&0.0& 0.9& 1.3&0.2& 0.6& 1.0\\
    \cmidrule(lr){2-20}
        & \tcell{MLP (MSE)} &1.0&3.9&7.4&0.2&2.0&3.2&0.2&1.0&2.0&0.0&0.2&0.7&0.0&0.2&0.5&0.0&0.0&0.0\\
        & \tcell{MLP (Contr.)} &35.9&58.0&63.9&19.4&39.3&45.7&1.2&3.2&5.2&2.0&5.9&10.8&2.0&5.7&11.8&1.0&2.5&4.7 \\
        & \gcell{Transformer (MSE)} &0.2&4.2&7.6&0.5&2.0&3.2&0.0&0.7&2.5&0.2&0.7&1.2&0.0&0.0&0.7&0.0&0.2&0.7 \\
        & \gcell{Transformer (Contr.)} &18.9&43.0&53.8&35.4&61.2&69.0&0.7&3.7&6.1&0.5&3.2&7.6&1.7&4.9&11.1&1.5&4.2&7.4 \\
             
        \midrule
        \parbox[t]{1em}{\multirow{8}{*}{\rotatebox[origin=c]{90}{\textbf{Flickr30K}}}}
        & Euclidean&48.2& 63.7& 68.7&47.7& 79.7& 88.8&0.0& 0.4& 1.0&0.3& 0.5& 1.1&0.1& 0.3& 1.0&0.1& 0.5& 1.0\\
        & Cosine&\textbf{70.1}& \textbf{89.1}& \textbf{93.3}&65.3& 87.9& 92.8&0.0& 0.4& 1.0&0.2& 0.5& 1.2&0.1& 0.6& 0.8&0.1& 0.6& 1.5\\
        & Manhattan&60.8& 81.9& 87.3&40.2& 70.8& 81.7&0.0& 0.5& 0.8&0.1& 0.5& 1.2&0.1& 0.5& 0.9&0.1& 0.4& 1.1\\
        & Chi-Square&0.0& 0.1& 0.4&0.1& 0.3& 0.6&0.0& 0.3& 0.4&0.0& 0.4& 1.2&0.0& 0.4& 0.9&0.0& 0.2& 0.7\\
    \cmidrule(lr){2-20}
        & \tcell{MLP (MSE)} &40.0&69.1&79.0&8.6&27.2&40.3&0.1&0.3&0.5&0.1&0.1&0.2&0.1&0.4&0.8&0.1&0.3&0.4 \\
        & \tcell{MLP (Contr.)} &42.6&72.0&81.4&33.3&64.3&75.0&0.2&0.9&2.0&3.3&11.8&19.0&5.4&20.3&29.6&1.6&6.5&11.7 \\
        & \gcell{Transformer (MSE)} &37.7&67.0&77.4&37.6&66.2&76.8&0.1&0.2&0.5&0.1&.01&0.2&0.1&0.5&0.7&0.0&0.1&0.4 \\
        & \gcell{Transformer (Contr.)} &29.4&61.6&73.9&34.0&63.5&75.0&0.5&1.8&3.1&3.6&12.1&20.2&6.2&21.6&31.8&3.4&13.5&20.5 \\
                 
        \midrule
        \parbox[t]{1em}{\multirow{8}{*}{\rotatebox[origin=c]{90}{\textbf{MS-COCO}}}}
        & Euclidean&32.2& 55.0& 65.2&48.0& 82.5& 93.4&0.2& 0.4& 0.9 &0.0& 0.2& 0.5&0.1& 0.5& 1.1&0.1& 0.5& 0.9\\
        & Cosine& 51.0& 82.3& 92.2&\textbf{64.7}& \textbf{90.1}& \textbf{96.0}&0.1& 0.3& 0.7&0.0& 0.2& 0.7&0.0& 0.3& 0.6&0.1& 0.6& 1.2\\
        & Manhattan&43.4& 73.5& 86.1&42.0& 74.9& 86.7&0.0& 0.5& 1.1&0.0& 0.3& 0.7&0.0& 0.7& 1.4&0.1& 0.5& 1.0\\
        & Chi-Square&0.0& 0.4& 1.0&0.0& 0.6& 1.0&0.1& 0.5& 1.0&0.1& 0.3& 0.9&0.0& 0.3& 0.8&0.1& 0.5& 1.3\\
    \cmidrule(lr){2-20}
        & \tcell{MLP (MSE)} &24.1&53.9&68.9&11.9&34.0&48.6&0.1&0.3&0.5&0.1&0.1&0.2&0.1&0.2&0.5&0.0&0.2&0.4 \\
        & \tcell{MLP (Contr.)} &27.2&57.7&71.5&31.7&63.9&77.2&0.2&1.2&1.9&3.5&14.7&24.5&4.9&19.0&30.1&1.9&9.1&16.0 \\
        & \gcell{Transformer (MSE)} &24.8&54.1&69.6&38.7&69.0&81.2&0.0&0.1&0.4&0.1&0.2&0.3&0.2&0.4&0.5&0.0&0.1&0.2 \\
        & \gcell{Transformer (Contr.)} &20.7&50.8&65.7&33.2&67.1&79.5&0.7&2.8&3.9&4.2&14.8&24.2&7.5&20.3&32.0&3.3&12.7&22.9 \\
             
        \bottomrule
    \end{tabular}
    }
    \label{tab:img2txt}
\end{table}

Furthermore, for unaligned representations extracted by Meta-Transformer and combined unimodal models, we observe that both MLPs and transformers with the custom contrastive loss outperform their counterparts trained with MSE loss across all the datasets. Out of all the trained models with different losses, transformers with custom contrastive loss achieve the best performance, except for representations extracted by ConViT/RoBERTa on IMDB, where MLPs defeat transformers with contrastive loss.

To better understand the impact of transformer design, 
we quantify the performance impact of the number of attention heads across all datasets in \autoref{tab:transformer_head}. 
Note that we do not add more cross-attention blocks, otherwise the model sizes increase drastically, easily causing overfitting on datasets.
From the table, we observe that the model performance varies on different datasets when using different number of heads.
Overall, transformer-based models with 4 heads produce greater benefits.

\begin{table}[htbp]
    \centering
    \caption{
        Retrieval performance (Recall@1) of transformer-based models with varying numbers of attention heads in $\{1, 2, 4\}$, trained using the custom contrastive loss on CLIP-extracted representations. All other settings are consistent with the experiments described in \autoref{sec:method}.
    }
    \begin{tabular}{l *6c}
    \toprule
         &\multicolumn{3}{c}{\textbf{Text-to-Image}}&\multicolumn{3}{c}{\textbf{Image-to-Text}}\\
         \cmidrule(lr){2-4}\cmidrule(lr){5-7}
         \textbf{Dataset}&1 &2 &4 &1 &2 &4 \\
    \midrule
         \textbf{IMDB}&19.4&18.9&18.9&20.4&20.4&18.9  \\
         \textbf{Flickr30K}&29.0&29.3&29.7&29.1&29.6&29.4  \\
         \textbf{MS-COCO}&20.4&21.7&22.0&19.8&19.8&20.7 \\
    \bottomrule
    \end{tabular}
    \label{tab:transformer_head}
\end{table}

From our experiments, we observe that the performance of neural networks is still falling behind cosine similarity. When using cosine metric,
CLIP outperforms in image-to-text tasks, achieving the best results on 2 out of 3 datasets. In contrast, BLIP performs better in text-to-image tasks, also winning on 2 out of 3 datasets, despite having the smallest modality gap described in \autoref{sec:embed_space}.

We believe this phenomenon is primarily related to differences in architecture and modality-specific representation learning. 
CLIP is optimized to produce global visual representations of entire images and strong image-text matching through symmetric contrastive training, which explains its advantage in image-to-text retrieval. 
In contrast, BLIP incorporates additional image-text interaction and language modeling objectives, enabling richer textual representations and stronger cross-modal semantic grounding, which can benefit text-to-image retrieval.
Plus, images contain more information than text, as text describes only a part of the image content, causing a partial semantic mismatch between modalities.
Importantly, the asymmetry is consistent with our geometric analysis. 
Although BLIP exhibits the smallest modality gap between image and text representations, a smaller modality gap does not necessarily translate into uniformly better performance across both retrieval directions.
Notably, BLIP consistently outperforms CLIP on MS-COCO.
These results highlight the significant variability in model performance across different datasets.
Moreover, cosine similarity cannot perform equally well for both tasks since the mean of cosine similarity across images with respect to all texts is not equal to the mean of cosine similarity across texts with respect to all images.
As for Euclidean and Manhattan distances, they perform similar to cosine similarity.
This is because both the Euclidean distance and Manhattan distance is inversely proportional to the cosine similarity of the normalized representations from CLIP and BLIP models.
For example, on a unit circle, a pair of normalized image and text representations with higher cosine similarity has a smaller angular separation, resulting in smaller Euclidean and Manhattan distances between them.


To investigate the impact of multiple captions, we use all 5 captions from the test split (1K) of Flickr30K 
to inspect the performance of CLIP and BLIP models based on cosine similarity. 
The results are shown in \autoref{tab:all_caps}.
For image-to-text retrieval tasks, we observe a performance decline as the retrieval scope widens (e.g., from \( P@1 \) to \( P@10 \)). This phenomenon arises because, in these experiments, each image can have up to five ground-truth captions. For text-to-image tasks, where each caption corresponds to one image, the upper bounds differ significantly.
For both retrieval tasks, the upper-bound precision values are constrained by the dataset structure (five captions per image). See \ref{app:precision_curve} for more details.

\begin{table}[htbp]
    \centering
    \caption{
        Precision ($\uparrow$) of retrieval tasks on the 1K test split of Flickr30K with 5 captions,
        using cosine similarity metric on representations extracted by CLIP and BLIP models.
    }
    \begin{tabular}{ccccccc}
        \toprule
         \multirow{2}*{}&\multicolumn{3}{c}{\textbf{Image-to-Text}}&\multicolumn{3}{c}{\textbf{Text-to-Image}}\\
         \cmidrule(lr){2-4}\cmidrule(lr){5-7}
         &P@1&P@5&P@10&P@1&P@5&P@10\\
        \midrule
        \textbf{CLIP}&79.30&58.06&36.27&58.92&16.69&9.01  \\
        \textbf{BLIP}&66.00&53.12&34.47&61.44&16.82&8.95\\
        \bottomrule
    \end{tabular}
    \label{tab:all_caps}
\end{table}

To conclude these experiments, 
we conducted statistical tests to corroborate the significance of our results. 
Pairwise comparisons (Bonferroni-Holm corrected) using the Chi-square test of proportions
revealed statistically significant differences
between CLIP/BLIP using cosine similarity and all the other models, in all datasets,
both in image-to-text and text-to-image retrieval experiments,
for all top-K results ($p < .001$).
No statistically significant differences were found
between Meta-Transformer, ResNet/BERT, ConViT/RoBERTa, and ConvNeXt/XLNet,
for any of the top-K results and any of the metrics ($p > .05$)
in any dataset or retrieval task.

\subsubsection{Sample Outputs}

Here we provide real-world application examples using representations from CLIP
to retrieve images given text queries and vice versa. 
The leftmost column depicts the input query (either images or texts), 
and the other columns show the top-5 retrieved results.

\begin{figure}[htbp]
    \centering
    \begin{subfigure}{\textwidth}
        \centering
        \includegraphics[width=\textwidth]{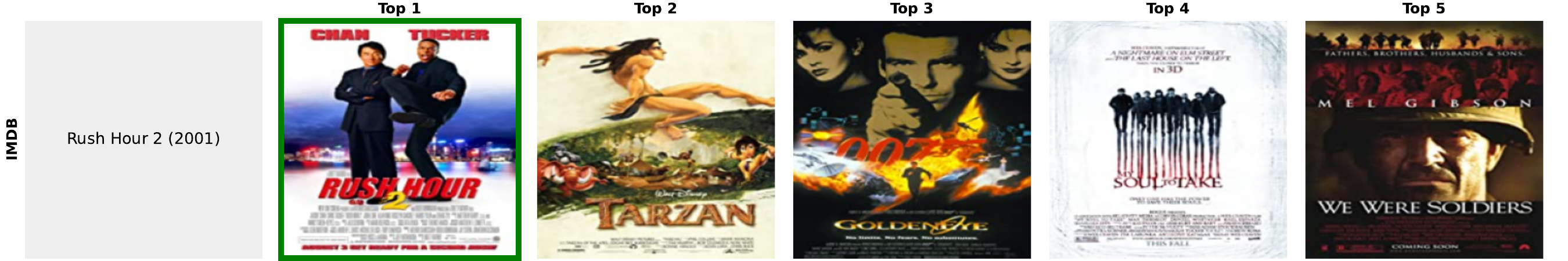}
    \end{subfigure}
    \hfill
    \begin{subfigure}{\textwidth}
        \centering
        \includegraphics[width=\textwidth]{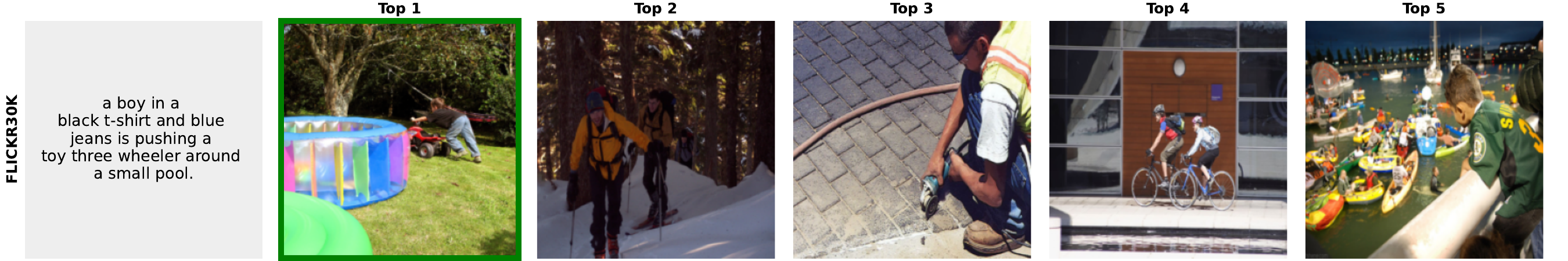}
    \end{subfigure}
    \hfill
    \begin{subfigure}{\textwidth}
        \centering
        \includegraphics[width=\textwidth]{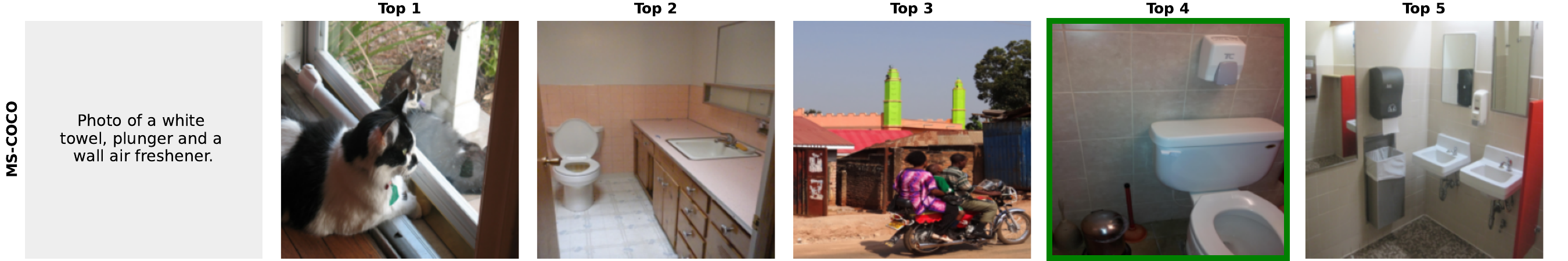}
    \end{subfigure}
    \caption{
        Image retrieval examples given randomly selected texts as input queries using CLIP. 
        \textcolor{ForestGreen}{Correct results} marked in green. 
        CLIP successfully retrieved the corresponding image on all datasets, 
        although for MS-COCO it was at position 4.
    }
    \label{fig:clip_img_retrieval}
\end{figure}

In \autoref{fig:clip_img_retrieval}, 
CLIP retrieved the right poster image for the given title as the top-1 result on IMDB. 
It also correctly retrieved the correct scenario as the top-1 result on Flickr30k. 
The top-4 result was retrieved for MS-COCO as well, 
although the dataset covers a wide range of complex and diverse scenarios.
From the examples, it might be surprising that CLIP performs so well on the IMDB dataset, 
since movie names are rather short and often not semantically related to their poster images. 
This might be the result of potential overlap between the training data of CLIP and the IMDB dataset,
as well as CLIP's excellent zero-shot performance. 
On the other hand, MS-COCO has more diverse images that cover more scenarios 
and is therefore more challenging for retrieval tasks, 
which underscores the lower performance compared to IMDB and Flickr30k, as in \autoref{tab:txt2img} and \autoref{tab:img2txt}. 

\begin{figure}[htbp]
    \centering
    \begin{subfigure}{\textwidth}
        \centering
        \includegraphics[width=\textwidth]{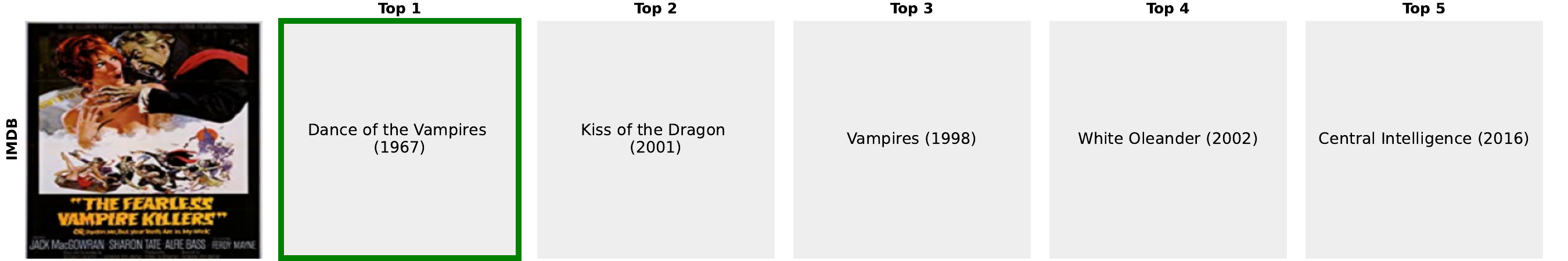}
    \end{subfigure}
    \hfill
    \begin{subfigure}{\textwidth}
        \centering
        \includegraphics[width=\textwidth]{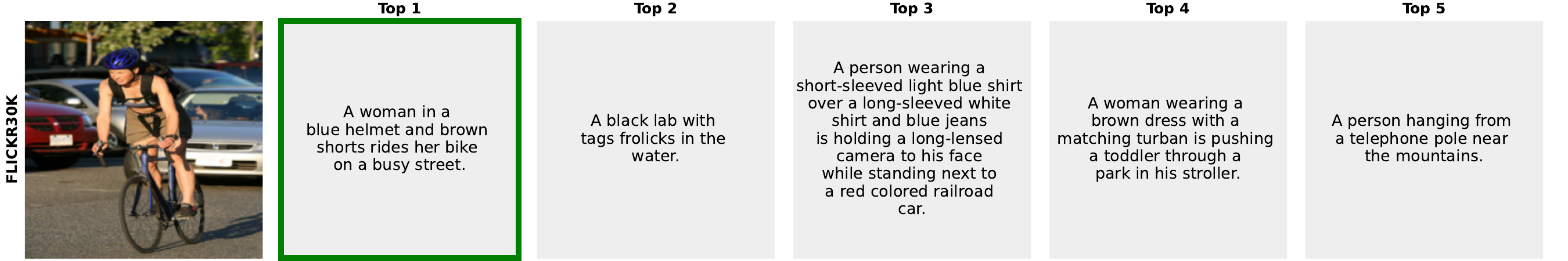}
    \end{subfigure}
    \hfill
    \begin{subfigure}{\textwidth}
        \centering
        \includegraphics[width=\textwidth]{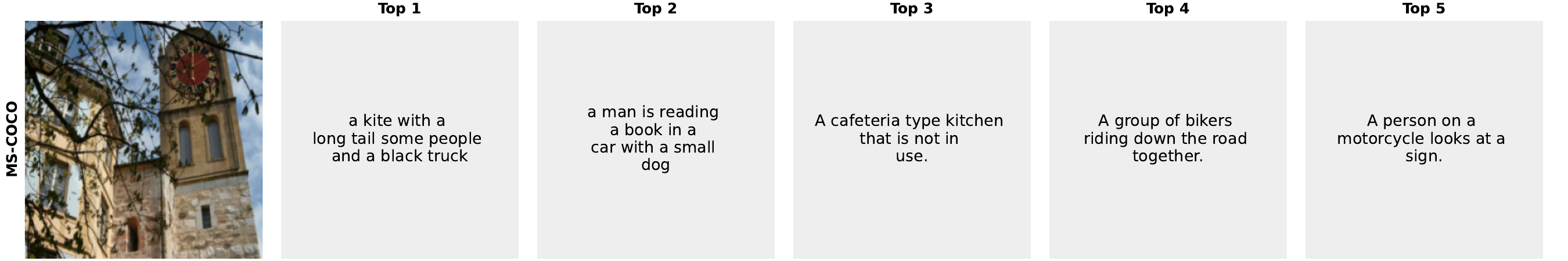}
    \end{subfigure}
    \caption{
        Text retrieval examples given randomly selected images as input queries using CLIP. 
        \textcolor{ForestGreen}{Correct results} marked in green. 
        CLIP successfully retrieved the corresponding text on IMDB and Flickr30K datasets, whereas it failed on MS-COCO.
    }
    \label{fig:clip_txt_retrieval}
\end{figure}

\autoref{fig:clip_txt_retrieval} shows the retrieval results of text descriptions when given images as input. 
CLIP successfully retrieved corresponding descriptions within top-5 results for IMDB and Flickr30K, 
whereas it failed on MS-COCO. 
From multiple rounds of retrieval results, 
we find that CLIP's language understanding capability is not symmetric to its visual understanding on certain cases, 
which could be a result of an asymmetric contrastive loss and different understanding capabilities of visual and language encoders.
Our retrieval results corroborate the aforementioned analysis.

The examples of retrieval results for BLIP and Meta-Transformer are shown in \ref{app:ext_example}. 
We noted that Meta-Transformer performed significantly worse than BLIP and CLIP models,
as informed by the results of the experiments presented in \autoref{sec:similarity} section, forming a narrow cone that is challenging to align either via existing metrics or neural networks,
even though Meta-Transformer's modality gap is close to those of CLIP and BLIP.  
\section{Discussion, Limitations, and Future Work}

Our experiments systematically assess the extent to which learned representations from multimodal and combined unimodal models can be leveraged in cross-modal retrieval tasks. To the best of our knowledge, this is the first work to extensively and systematically investigate this underexplored direction.
The key insights from our experiments are:
\begin{itemize}
    \item The geometry of multimodal representations does not necessarily correlate with retrieval performance, and the smaller modality gap (either quantified by Euclidean or distribution distance) does not reflect better performance.
    \item When both textual and visual embeddings are extracted from the same VLM, cosine similarity consistently outperforms alternative metrics, and the learned ones via different neural networks with different losses.
    \item The proposed custom contrastive loss to align separate textual and visual representations proves more effective than MSE overall for both transformer-based models and MLPs, and the predicted scores are more discriminative than cosine similarity, as evidenced by \ref{app:scores}.
\end{itemize}

To motivate future research directions, we outline the key limitations and open questions emerging from our findings.
First of all, our experiments show that there is no single combination of model and loss function that consistently outperforms the rest in all tasks and datasets. 
For example, using cosine similarity, CLIP outperforms BLIP on the IMDB dataset for both text-to-image and image-to-text retrieval tasks, whereas BLIP outperforms CLIP on the MS-COCO dataset for both tasks.
The causes can be the dataset-specific characteristics (e.g., diversity), and the different training objectives used for each of them.

Furthermore, the consistency between Wasserstein distance in \autoref{tab:wasserstein_dist} and modality-gap measurements in \autoref{fig:tsne_embeddings} suggests that future alignment methods may benefit from explicitly minimizing discrepancies between image and text distributions. Unlike point-wise objectives, distribution-aware objectives could preserve global geometry while reducing cross-modal discrepancies. 
In addition, applying t-SNE to visualize representations extracted from different layers of a neural network 
may yield varying results. 
This variability arises due to evolving representations across network layers, 
which can lead to changes in the underlying data distribution~\cite{distributionshifts}. 
When the input distribution to t-SNE is altered, the resulting low-dimensional embeddings reflect 
the distribution of the transformed representations, rather than the original data distribution. 
Consequently, dimensionality reduction techniques such as t-SNE primarily reveal feature-space structures 
induced by the model, rather than faithfully representing the intrinsic structure of the raw data.

%

In addition, computational scalability presents another significant constraint: 
Cross-modal similarity computation scales quadratically, $\mathcal{O}(N^2)$ with dataset cardinality $N$ 
under one-to-one modality mapping assumptions. 
This complexity imposes practical limitations on large-scale retrieval systems.

Each encoder of pre-trained multimodal models has its own limitations as well. On the one hand, current pre-trained image encoders exhibit limited capability in capturing fine-grained visual details, 
deterring precise image-text alignment. 
However, we emphasize that, as mentioned in \autoref{sec:datasets}, this limitation is unrelated to image resizing, 
as evidenced by SPARC~\cite{sparc} which achieves fine-detail understanding 
on $224 \times 224$ pixel inputs through contrastive language-visual alignment. 
While models like the one presented by \citet{retrievalenhancedclip} demonstrate promising cross-modal refinement, 
their restricted availability hinders practical adoption.
On the other hand, pre-trained text encoders face two main limitations: 
(1)~Input truncation (e.g., 76 tokens in CLIP, 512 in BLIP) constrains comprehension of lengthy documents, 
potentially addressable through LLM-based summarization preprocessing; 
(2)~Performance degradation on non-English texts due to English-centric training, 
exacerbating challenges for low-resource languages~\cite{engpivotal}. 
Effective multilingual support necessitates comprehensive training data in target languages 
to capture language-specific semantic nuances.

Finally, we observed a geometric constraint in multimodal embedding spaces: the modality gap between representations  exacerbated by contrastive learning objectives. 
This fundamental limitation warrants novel geometric alignment techniques at the distributional level in future multimodal architectures.
Taken together, these findings motivate a geometric alignment framework that simultaneously optimizes local semantic alignment between paired samples and global distribution alignment across modalities while preserving embedding diversity.
We believe that future retrieval models incorporating these three objectives through optimal transport, distribution matching, or isotropy-preserving regularization may overcome the identified limitations.
\section{Conclusion}

This work systematically examines the alignment of visual and textual representations through both multimodal models and combined unimodal models, employing four established metrics alongside two different architectures with two loss functions that learn to align representations. 
Our analysis demonstrates that embedding space alignment can be effectively quantified through distributional distance metrics, 
which provide a complementary perspective for geometric characteristics of embedding spaces. 
Crucially, we observe that VLMs trained with contrastive loss, despite inducing a modality gap, 
consistently outperform combined unimodal counterparts when evaluated using cosine similarity. 
Furthermore, we establish that contrastive loss is more effective for modeling the complex cross-modal interactions between representations
for both MLPs and transformer-based models, especially for the unaligned representations extracted by respective unimodal models. With our proposed custom objective function, we resolve the discriminativeness caused by cosine similarity from pre-trained VLMs.
Collectively, these findings provide substantive new research perspectives for cross-modal information retrieval utilizing multimodal frameworks, inspiring the future research on representation alignment.
\section*{Acknowledgments}
We thank Niculae Sebe and Fabrizio Silvestri for reviewing an earlier version of this paper. The authors also acknowledge Luxembourg’s national supercomputer MeluXina and the LuxProvide management team for their support. This research was supported by the PRIDE programme of the Luxembourg National Research Fund (FNR) through the Doctoral Training Unit “Deep Data Science of Digital History” (FNR PRIDE21/16758026/D4H), and by the European Innovation Council through the Pathfinder programme (SYMBIOTIK, grant 101071147).





\small
\bibliographystyle{elsarticle-num-names} 
\bibliography{ref}
\normalsize

\renewcommand{\thefigure}{\Alph{section}.\arabic{figure}} 
\renewcommand{\thetable}{\Alph{section}.\arabic{table}} 
\counterwithin{figure}{section} 
\counterwithin{table}{section}
\appendix
\section{Model Parameters}
\label{app:params}

We use MLPs and transformer-based neural networks to align the representations extracted from the respective encoders considered in our experiments. Since the embedding dimensions produced by different encoders vary, we adjust the input size of each network accordingly, which leads to different model sizes, as shown in \autoref{tab:params}.

\begin{table}[htbp]
    \centering
    \caption{
        Number of model parameters when using representations extracted by different encoders. For brevity, we omit the repeated sizes with the ditto mark (\texttt{"}), since the trained model size is only affected by input dimensions rather than datasets. In our case, we have inputs of dimension 512 and 768, respectively.
    }
    \resizebox{\linewidth}{!}{
    \begin{tabular}{l *6c}
    \toprule
         &\textbf{CLIP} &\textbf{BLIP} &\textbf{Meta-Transf.} &\textbf{ResNet/BERT} &\textbf{ConViT/RoBERTa} &\textbf{ConvNeXt/XLNet}  \\
    \midrule
         Embedding dim.&512&512&768&512&768&768  \\
         MLP size&656,385&\texttt{"}&918,529&\texttt{"}&\texttt{"}&\texttt{"}  \\
         Transformer size&560,897&\texttt{"}&626,433&\texttt{"}&\texttt{"}&\texttt{"}  \\
    \bottomrule
    \end{tabular}
    }
    \label{tab:params}
\end{table}
\section{Model Predicted Scores}
\label{app:scores}

To provide further insights of the discriminativeness of the proposed custom contrastive loss function, we also show the average scores predicted by trained models with both MSE and the proposed loss for positive pairs and negative ones in \autoref{tab:scores_txt2img} and \autoref{tab:scores_img2txt}. 
From both retrieval tasks, we observe that the custom contrastive loss can effectively produce positive scores for positive pairs and negative scores for negative pairs, which addresses the positive-scores-for-all-pairs problem using cosine similarity on well-aligned representations extracted from CLIP and BLIP.
In some cases where scores for both positive pairs and negative pairs are negative, however, the differences between them are sufficiently large to discriminate positives from negatives.

\begin{table}[htbp]
    \centering
    \caption{
    Average predicted scores via trained neural networks in \autoref{tab:txt2img} with different losses on different datasets when using varying feature extractors for \textit{text-to-image} retrieval tasks. The first number is the average score of positive pairs, while the second number is the average score of negative pairs.
    }
    \resizebox{\linewidth}{!}{
    \begin{tabular}{ll *6c}
    \toprule
         &\textbf{Model}&\textbf{CLIP}&\textbf{BLIP}&\textbf{Meta-Transf.}&\textbf{ResNet/BERT}&\textbf{ConViT/RoBERTa}&\textbf{ConvNeXt/XLNet}\\
    \midrule
         \multirow{4}{*}{\rotatebox[origin=c]{90}{\tiny\textbf{IMDB}}}
         &MLP (MSE)&0.20, 0.20&0.23, 0.23&1.0, 1.0&-0.03, -0.03&-0.02, -0.02&-0.02, -0.02  \\ 
         &MLP (Contr.)&0.30, -0.92&0.22, -0.37&0.05, -0.08&-0.04, -0.35&-0.08, -0.55&0.02, -0.03  \\ 
         &Transformer (MSE)&0.22, 0.20&0.24, 0.24&0.97, 0.97&-0.02, -0.02&-0.02, -0.02&-0.02, -0.02  \\ 
         &Transformer (Contr.)&0.30, -0.91&0.57, -0.83&0.09, -0.14&0.00, -0.49&0.32, -0.58&0.04, -0.38  \\ 
    \midrule
        \multirow{4}{*}{\rotatebox[origin=c]{90}{\tiny\textbf{Flickr30K}}}
         &MLP (MSE)&0.29, 0.16&0.34, 0.19&1.0, 1.0&-0.02, -0.02&0.01, 0.01&-0.01, -0.01  \\ 
         &MLP (Contr.)&0.72, -0.97&0.78, -0.90&0.10, -0.13&0.26, -0.82&0.42, -0.86&0.41, -0.60  \\ 
         &Transformer (MSE)&0.31, 0.17&0.43, 0.20&0.98, 0.98&-0.02, -0.02&0.01, 0.01&-0.01, -0.01  \\ 
         &Transformer (Contr.)&0.78, -0.96&0.83, -0.95&0.16, -0.24&0.47, -0.80&0.53, -0.87&0.44, -0.83  \\ 
    \midrule
         \multirow{4}{*}{\rotatebox[origin=c]{90}{\tiny\textbf{MS-COCO}}}
         &MLP (MSE)&0.28, 0.15&0.35, 0.20&1.0, 1.0&-0.02, -0.01&0.01, 0.01&0.01, 0.01  \\ 
         &MLP (Contr.)&0.75, -0.97&0.83, -0.94&0.16, -0.18&0.45, -0.87&0.53, 0.90&0.55, -0.75  \\ 
         &Transformer (MSE)&0.30, 0.15&0.44, 0.20&0.98, 0.98&-0.02, -0.01&0.01, 0.01&-0.01, -0.01  \\ 
         &Transformer (Contr.)&0.81, -0.96&0.88, -0.97&0.25, -0.32&0.56, -0.85&0.65, -0.90&0.55, -0.88  \\ 
    \bottomrule
    \end{tabular}
    }
    \label{tab:scores_txt2img}
\end{table}

\begin{table}[htbp]
    \centering
    \caption{
    Average predicted scores via trained neural networks in \autoref{tab:img2txt} with different losses on different datasets when using varying feature extractors for \textit{image-to-text} retrieval tasks. The first number is the average score of positive pairs, while the second number is the average score of negative pairs.
    }
    \resizebox{\linewidth}{!}{
    \begin{tabular}{ll *6c}
    \toprule
         &\textbf{Model}&\textbf{CLIP}&\textbf{BLIP}&\textbf{Meta-Transf.}&\textbf{ResNet/BERT}&\textbf{ConViT/RoBERTa}&\textbf{ConvNeXt/XLNet}  \\
    \midrule
         \multirow{4}{*}{\rotatebox[origin=c]{90}{\tiny\textbf{IMDB}}}
         &MLP (MSE)&0.21, 0.20&0.23, 0.23&1.0, 1.0&-0.02, -0.02&-0.02, -0.02&-0.02, -0.02  \\ 
         &MLP (Contr.)&0.30, -0.92&0.23, -0.38&0.04, -0.07&-0.07, -0.39&-0.07, -0.51&0.03, -0.02  \\ 
         &Transformer (MSE)&0.22, 0.20&0.24, 0.24&0.97, 0.97&-0.21, -0.22&-0.02, -0.02&-0.02, -0.02  \\ 
         &Transformer (Contr.)&0.28, -0.91&0.56, -0.83&0.10, -0.15&0.00, -0.49&0.00, -0.60&0.05, -0.39  \\ 
    \midrule
        \multirow{4}{*}{\rotatebox[origin=c]{90}{\tiny\textbf{Flickr30K}}}
         &MLP (MSE)&0.30, 0.16&0.33, 0.19&1.0, 1.0&-0.02, -0.02&0.01, 0.01&-0.01, -0.01  \\ 
         &MLP (Contr.)&0.71, -0.98&0.79, -0.80&0.10, -0.13&0.26, -0.82&0.39, -0.87&0.40, -0.61  \\ 
         &Transformer (MSE)&0.31, 0.17&0.42, 0.20&0.98, 0.98&-0.02, -0.02&0.01, 0.01&-0.01, -0.01  \\ 
         &Transformer (Contr.)&0.77, -0.96&0.82, -0.95&0.17, -0.25&0.47, -0.79&0.53, -0.87&0.45, -0.82  \\ 
    \midrule
         \multirow{4}{*}{\rotatebox[origin=c]{90}{\tiny\textbf{MS-COCO}}}
         &MLP (MSE)&0.28, 0.15&0.34, 0.20&1.0, 1.0&-0.02, -0.01&0.01, 0.01&-0.01, -0.01  \\ 
         &MLP (Contr.)&0.75, -0.97&0.83, -0.94&0.16, -0.18&0.44, -0.87&0.52, -0.91&0.54, -0.75  \\ 
         &Transformer (MSE)&0.29, 0.15&0.44, 0.21&0.98, 0.98&-0.02, -0.01&0.01, 0.01&-0.01, -0.01  \\ 
         &Transformer (Contr.)&0.81, -0.96&0.89, -0.97&0.25, -0.32&0.59, -0.85&0.64, -0.90&0.55, -0.87  \\ 
    \bottomrule
    \end{tabular}
    }
    \label{tab:scores_img2txt}
\end{table}
\section{Precision for images with 5 captions from Flickr30K}
\label{app:precision_curve}

\begin{figure}[htbp]
    \centering
    \includegraphics[width=0.5\linewidth]{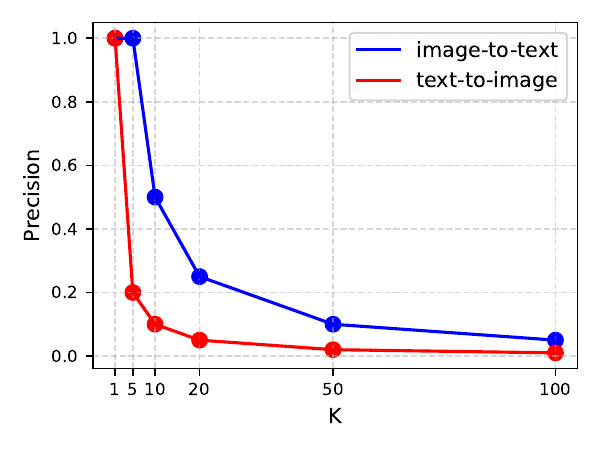}
    \caption{Upper bounds of P@K on the test split of Flickr30K. The drop is more significant for text-to-image retrieval tasks, as each annotation only corresponds to one image, whereas each image matches five annotations for image-to-text retrieval instead.}
    \label{fig:precision_decline}
\end{figure}

\autoref{fig:precision_decline} shows the upper bounds for retrieval Precision at different ranking sizes $K$.
As \( K \) increases, the denominator grows faster than the numerator. While five captions per image increase the pool of potentially relevant matches, this advantage diminishes when \( K > 5 \), as the fixed number of ground-truth captions cannot offset the quadratic growth of retrieved samples.
Here, Precision drops more sharply because the single relevant image per caption cannot compensate for the rapidly growing denominator. This asymmetry highlights how dataset design (multiple captions per image vs. single-image matches) fundamentally impacts P@K trends in cross-modal retrieval. Notably, at \( K=5 \), image-to-text retrieval maintains perfect Precision (\( 1.0 \)) while text-to-image precision collapses to \( 0.2 \), demonstrating the critical role of annotation multiplicity in retrieval effectiveness.

\section{BLIP and Meta-Transformer Retrieval Examples}\label{app:ext_example}

\begin{figure}[!t]
    \centering
    \begin{subfigure}{\textwidth}
        \centering
        \includegraphics[width=\textwidth]{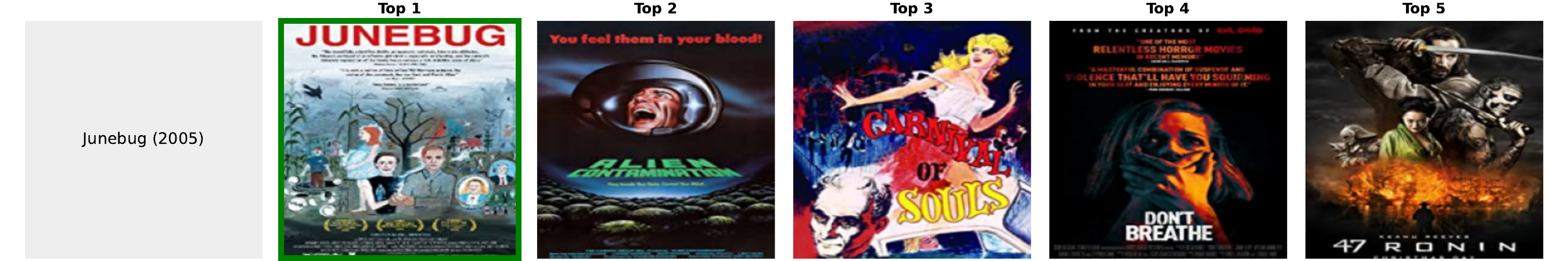}
        \caption{IMDB}
        \label{subfig:blip_yes_imdb}
    \end{subfigure}
    \hfill
    \begin{subfigure}{\textwidth}
        \centering
        \includegraphics[width=\textwidth]{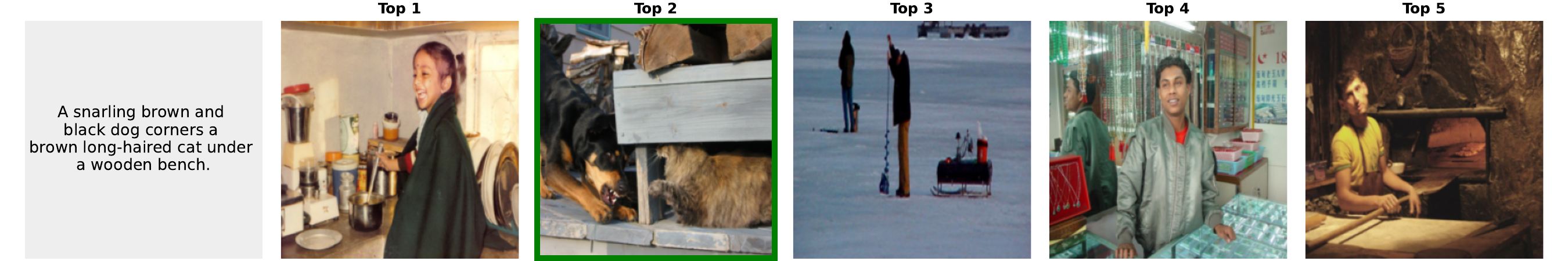}
        \caption{Flickr30K}
        \label{subfig:blip_yes_flickr30k}
    \end{subfigure}
    \hfill
    \begin{subfigure}{\textwidth}
        \centering
        \includegraphics[width=\textwidth]{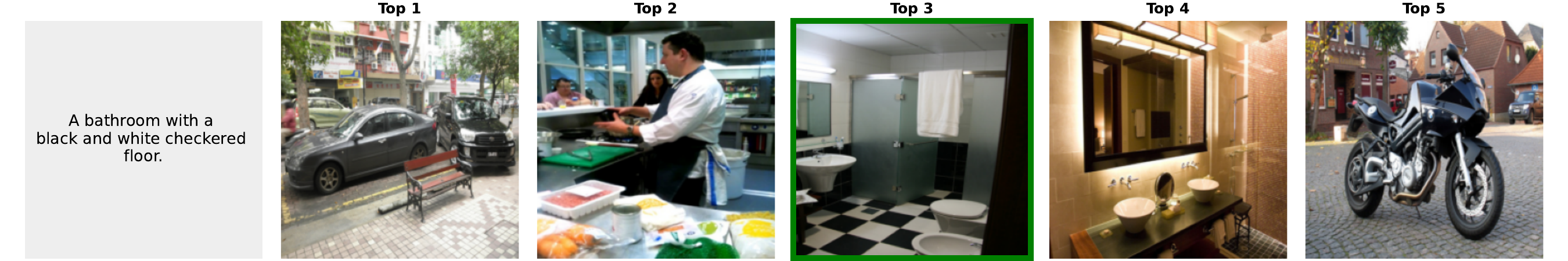}
        \caption{MS-COCO}
        \label{subfig:blip_yes_coco}
    \end{subfigure}
    \caption{Image retrieval examples given randomly selected texts as input queries using BLIP on three datasets. Correct results are marked in \textcolor{ForestGreen}{green}. BLIP successfully retrieves corresponding images on all datasets.}
    \label{fig:blip_img_retrieval}
\end{figure}

\begin{figure}[!t]
    \centering
    \begin{subfigure}{\textwidth}
        \centering
        \includegraphics[width=\textwidth]{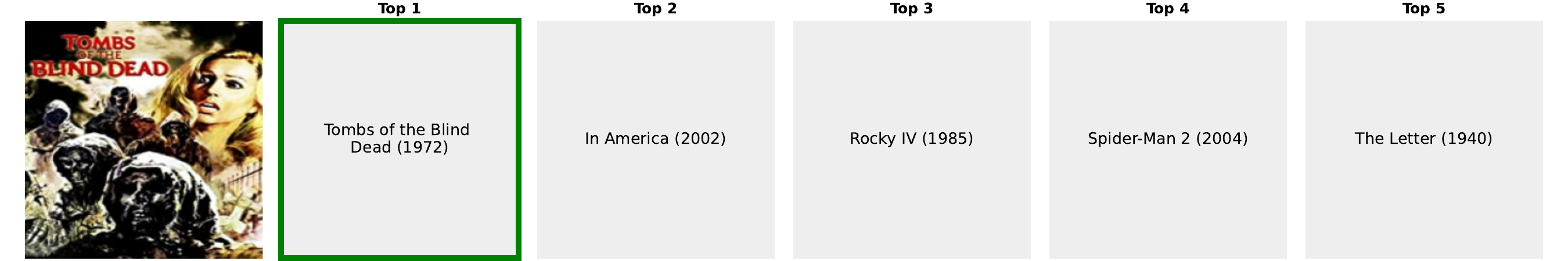}
        \caption{IMDB}
        \label{subfig:blip_no_imdb}
    \end{subfigure}
    \hfill
    \begin{subfigure}{\textwidth}
        \centering
        \includegraphics[width=\textwidth]{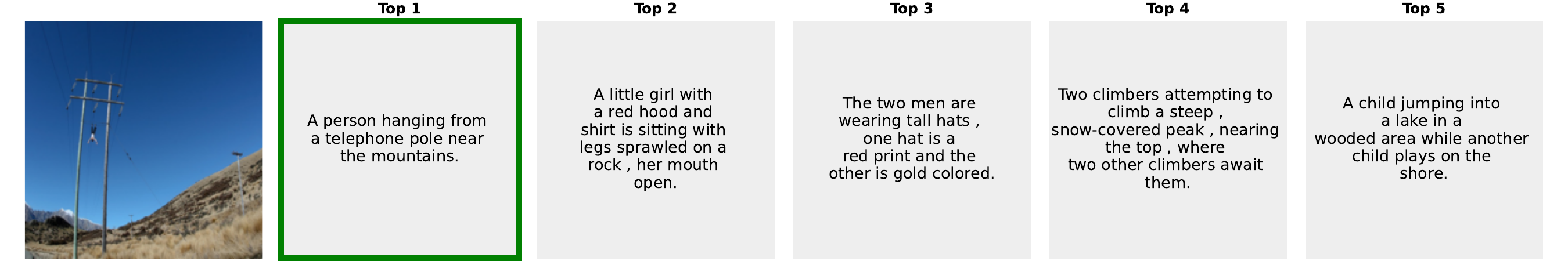}
        \caption{Flickr30K}
        \label{subfig:blip_no_flickr30k}
    \end{subfigure}
    \hfill
    \begin{subfigure}{\textwidth}
        \centering
        \includegraphics[width=\textwidth]{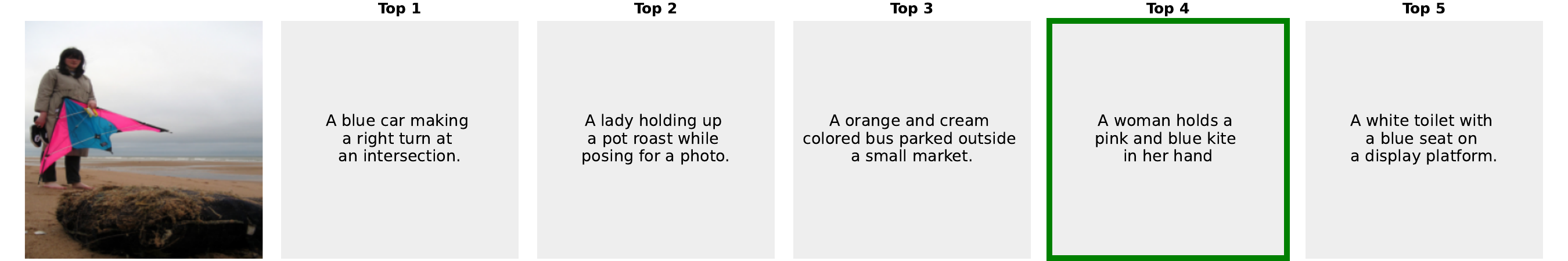}
        \caption{MS-COCO}
        \label{subfig:blip_no_coco}
    \end{subfigure}
    \caption{Text
    retrieval examples given randomly selected images as input queries using BLIP on three datasets. Correct results are marked in \textcolor{ForestGreen}{green}. BLIP successfully retrieves corresponding texts on all datasets.}
    \label{fig:blip_txt_retrieval}
\end{figure}

We show more retrieval examples on text-to-image and image-to-text tasks using BLIP model in \autoref{fig:blip_img_retrieval} and \autoref{fig:blip_txt_retrieval}. From both of the figures, we observe a good retrieval performance on all 3 datasets when using BLIP model.

\begin{figure}[!t]
    \centering
    \begin{subfigure}{\textwidth}
        \centering
        \includegraphics[width=\textwidth]{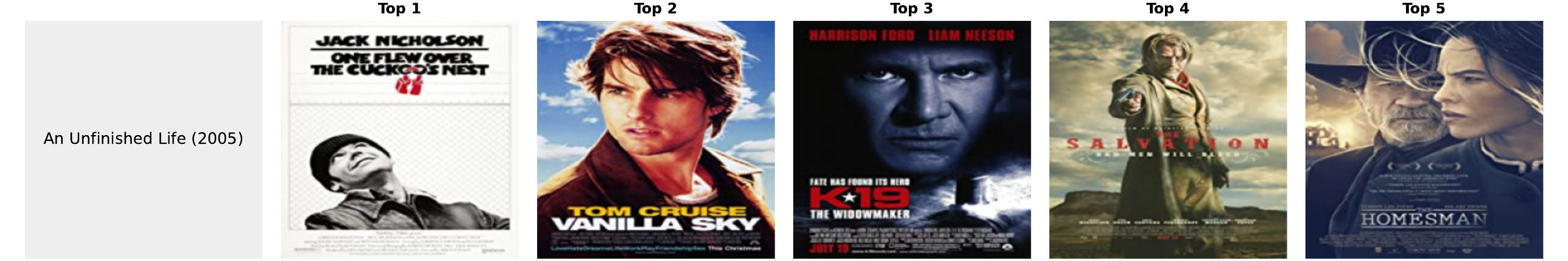}
        \caption{IMDB}
        \label{subfig:meta_yes_imdb}
    \end{subfigure}
    \hfill
    \begin{subfigure}{\textwidth}
        \centering
        \includegraphics[width=\textwidth]{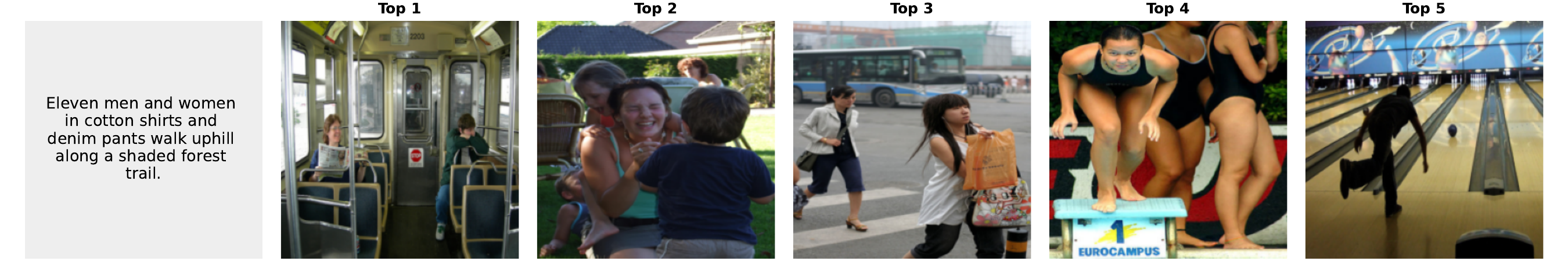}
        \caption{Flickr30K}
        \label{subfig:meta_yes_flickr30k}
    \end{subfigure}
    \hfill
    \begin{subfigure}{\textwidth}
        \centering
        \includegraphics[width=\textwidth]{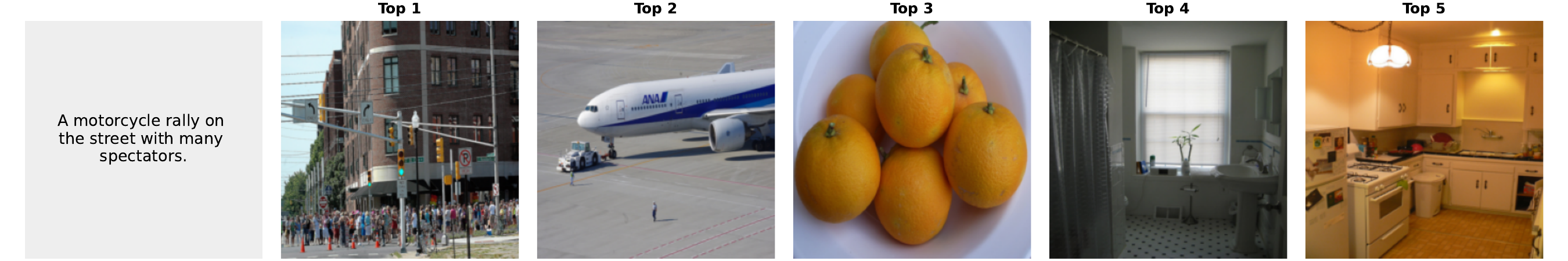}
        \caption{MS-COCO}
        \label{subfig:meta_yes_coco}
    \end{subfigure}
    \caption{Image retrieval examples given randomly selected texts as input queries using Meta-Transformer on three datasets. No correct results are retrieved in these cases.}
    \label{fig:meta_img_retrieval}
\end{figure}

\begin{figure}[!t]
    \centering
    \begin{subfigure}{\textwidth}
        \centering
        \includegraphics[width=\textwidth]{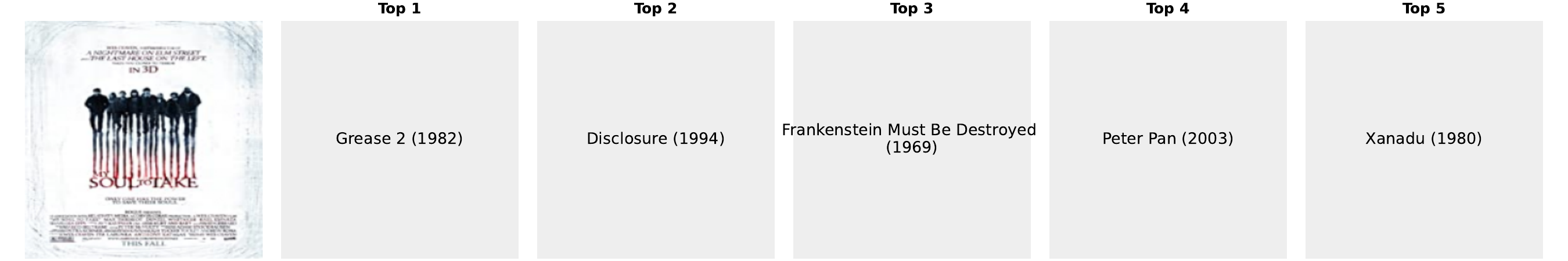}
        \caption{IMDB}
        \label{subfig:meta_no_imdb}
    \end{subfigure}
    \hfill
    \begin{subfigure}{\textwidth}
        \centering
        \includegraphics[width=\textwidth]{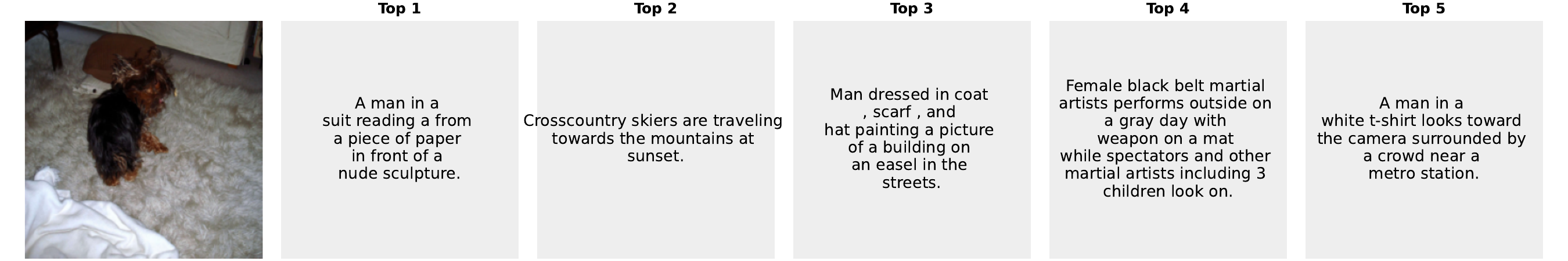}
        \caption{Flickr30K}
        \label{subfig:meta_no_flickr30k}
    \end{subfigure}
    \hfill
    \begin{subfigure}{\textwidth}
        \centering
        \includegraphics[width=\textwidth]{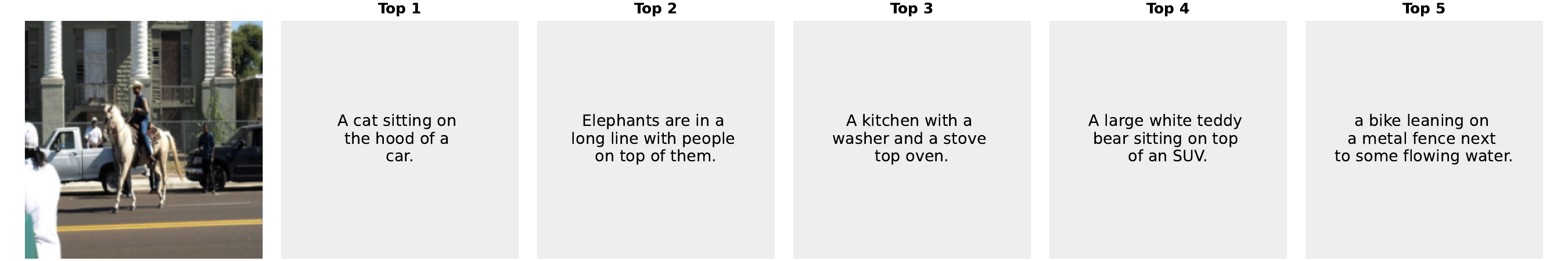}
        \caption{MS-COCO}
        \label{subfig:meta_no_coco}
    \end{subfigure}
    \caption{Text retrieval examples given randomly selected images as input queries using Meta-Transformer on three datasets. No correct results are retrieved in these cases.}
    \label{fig:meta_txt_retrieval}
\end{figure}

Representations extracted by Meta-Transformer perform poorly when used to match texts with images, as no results are correctly retrieved. Results are shown in \autoref{fig:meta_img_retrieval} and \autoref{fig:meta_txt_retrieval}, from which we barely observe connections between input queries and outputs.

\end{document}